\documentclass[prb,twocolumn,aps,showpacs,footinbib,10pt,numerical,superscriptaddress]{revtex4-1}
\usepackage[english]{babel}
\usepackage[utf8]{inputenc}
\usepackage{amsmath}
\usepackage{amsfonts}
\usepackage{amssymb}
\usepackage{graphicx}
\usepackage{lmodern}
\usepackage{color}
\usepackage{enumerate}
\usepackage{mathtools}
\usepackage{hyperref}
\usepackage[normalem]{ulem}
\usepackage{bbm}
\usepackage{bbold}
\usepackage{float}
\usepackage[caption=false]{subfig}
\usepackage{enumitem}
\usepackage{verbatim}
\usepackage[]{placeins}
\usepackage{lipsum}
\RequirePackage[normalem]{ulem} 
\RequirePackage{color}\definecolor{RED}{rgb}{1,0,0}\definecolor{BLUE}{rgb}{0,0,1} 
\providecommand{\DIFaddend}{} 

\begin{document}

\title{Voltage-tunable Majorana bound states in time-reversal symmetric bilayer quantum spin Hall hybrid systems}

\author{F.~Schulz}
\affiliation{Department of Physics, University of Basel, CH-4056 Basel, Switzerland}
\author{ J.~C.~Budich}
\affiliation{Institute of Theoretical Physics, Technical University of Dresden, D-01062 Dresden, Germany}
\author{ E.~G.~Novik}
\affiliation{Institute of Theoretical Physics, Technical University of Dresden, D-01062 Dresden, Germany}
\author{P.~Recher}
\affiliation{Institute of Mathematical Physics, Technical University of Braunschweig, D-38106 Braunschweig, Germany}
\affiliation{Laboratory for Emerging Nanometrology Braunschweig, D-38106 Braunschweig, Germany}
\author{B.~Trauzettel}
\affiliation{Institute of Theoretical Physics and Astrophysics, University of W\"urzburg, D-97074 W\"urzburg, Germany}
\affiliation{W\"urzburg-Dresden Cluster of Excellence ct.qmat, Germany}

\date{\today}

\begin{abstract}
{We investigate hybrid structures based on a bilayer quantum spin Hall
system in proximity to an $s$-wave superconductor as a platform to mimic time-reversal symmetric topological superconductors. In this bilayer setup, the induced pairing can be of intra- or inter-layer type, and domain walls of those
different types of pairing potentials host Kramers partners (time-reversal conjugate pairs) of
Majorana bound states. Interestingly, we discover that such topological interfaces providing Majorana bound states can also be achieved in an otherwise homogeneous system by a spatially dependent inter-layer gate voltage. This gate voltage causes the relative electron densities of the two layers to vary accordingly which suppresses the inter-layer pairing in regions with strong gate voltage. We identify particular transport signatures (zero-bias anomalies) in a five-terminal setup that are clearly related to the presence of Kramers pairs of Majorana bound states.
\vspace{1cm}}\\
\vspace{1cm}
\end{abstract}
\maketitle

\section{Introduction}\label{seq.:sectionI}
One of the fundamental prerequisites of topological quantum computing \cite{dassarma} (TQC) is based to the hardware exhibiting quasi-particles with non-Abelian exchange statistics. The primary example of such non-Abelian quasi-particles is provided by Majorana bound states (MBS). They are zero energy excitations with anyonic braiding properties. There is a plethora of theoretical proposals \cite{Read2000,kitaev2001,fu2008,fu2009,dassarma2010,oppen2010,alicea2010,alicea2012,beenakker2013}, predicting favorable conditions for hosting MBS. Moreover, experimental evidence of MBS has been reported in several systems \cite{kouwen2012,Das2012,marcus2016,marcus22016,Zhang2018}.

In particular, superconducting hybrid structures have been proposed to host MBS. Those hybrids can be realized in various ways, for instance, by coupling topological insulators (TIs) \cite{Kane2009b,beenakker2009,trauz2014,trauz22014,recher2015,keidel2018}, semiconducting nanowires \cite{dassarma2010,oppen2010,lossklino2012,lee2012,aguado2012,lossklino2013,simon2013}, or chains of adatoms \cite{bernevig2013,loss2013,mcdonald2014,yazdani2014,franke2015,meyer2016}  to an ordinary $s$-wave superconductor (SC) via the proximity effect. For our proposal, it is important to be able to attain a decent proximity effect between a quantum spin Hall (QSH) insulator -- a two-dimensional (2D) TI -- and an $s$-wave SC. Fortunately, this task has already been achieved in the laboratory in QSH insulators based on HgTe in proximity to Al or Nb \cite{molenkamp2014,kouwen2015,yacoby2017,molenkamp2017,Deacon2017}.

Inspired by the work of Klinovaja, Yacoby, and Loss \cite{Klino2014}, we investigate bilayer QSH systems in proximity to an ordinary $s$-wave SC (see Fig.~\ref{fig.:fullsetup} for a schematic). In this setup, we assume competing SC pairings, namely the direct pairing within the individual layers and the crossed pairing between the layers. The latter pairing can be interpreted as a Cooper pair splitting between the layers \cite{recher2002,fisher2002,Sato2010,recher2013}. Importantly, a significant inter-layer pairing as compared to the intra-layer pairing can only be realized in the presence of interactions. This requirement can be viewed as a consequence of a deeper theoretical concept, that interactions are necessary for the development of time-reversal symmetric topological superconductivity in low spatial dimensions \cite{flensberg2016a,flensberg2016b, haim2018}.

It has been shown before \cite{Klino2014} that Kramers pairs of MBS are located at the boundaries between two regions in space where the two different pairings (direct and crossed) dominate, respectively. Remarkably, we demonstrate below how an additional inter-layer gate voltage can provide an experimental knob for locally tuning the domain boundary of the superconducting pairing potentials. This approach works because the inter-layer gate voltage effectively suppresses the crossed pairing.

We present analytical results based on an effective model of helical edge states with proximity induced pairing, and corroborate our findings by numerical data on a microscopic lattice model of the 2D bilayer system. We start by investigating the spectral properties of the hybrid system. Based on the two complementary models, we examine the emergence of MBS at a spatial interface between two inequivalent Bogoliubov-de-Gennes (BdG) band structures.

Subsequently, we consider a five-terminal setup which allows us to distinguish all transport processes (electron reflection, Andreev reflection, electron cotunneling, and crossed Andreev reflection) by multi-terminal transport characteristics. Calculating the conductance matrix following the approaches by Blonder, Tinkham, and Klapwijk (BTK) \cite{BTK} as well as B\"uttiker \cite{buettiker}, we find -- in the presence of MBS due to a domain wall in the superconducting pairing -- dips (in the local conductance) and peaks (in the nonlocal conductance). By tuning the inter-layer gate voltage, this picture can even be inverted, i.e. we can find peaks for the local and dips for the nonlocal conductance. We carefully explain below to which extent the zero-bias features in the conductance are indeed related to the emergence of Kramers pairs of MBS.

The article is organized as follows: In Sec.~\ref{sec.:sectionII}, we present the full model in a tight-binding and effective low energy Hamiltonian description. In Sec.~\ref{sec.:sectionIII}, we examine the spectral properties of this model, where we show two different gap closing mechanisms. In Sec.~\ref{sec.:sectionIV}, we explain the appearance of MBS on a spatial interface between two different band gaps in the BdG picture. In Sec.~\ref{sec.:sectionV}, we present zero bias anomalies in local and non-local differential conductances. Finally, we conclude in Sec.~\ref{sec.:sectionVI}. Some technical details are moved to App.~\ref{sec.:appA} and App.~\ref{sec.:appB}.

\section{Model}\label{sec.:sectionII}

\begin{figure}
\centering
\includegraphics[width=0.9\linewidth]{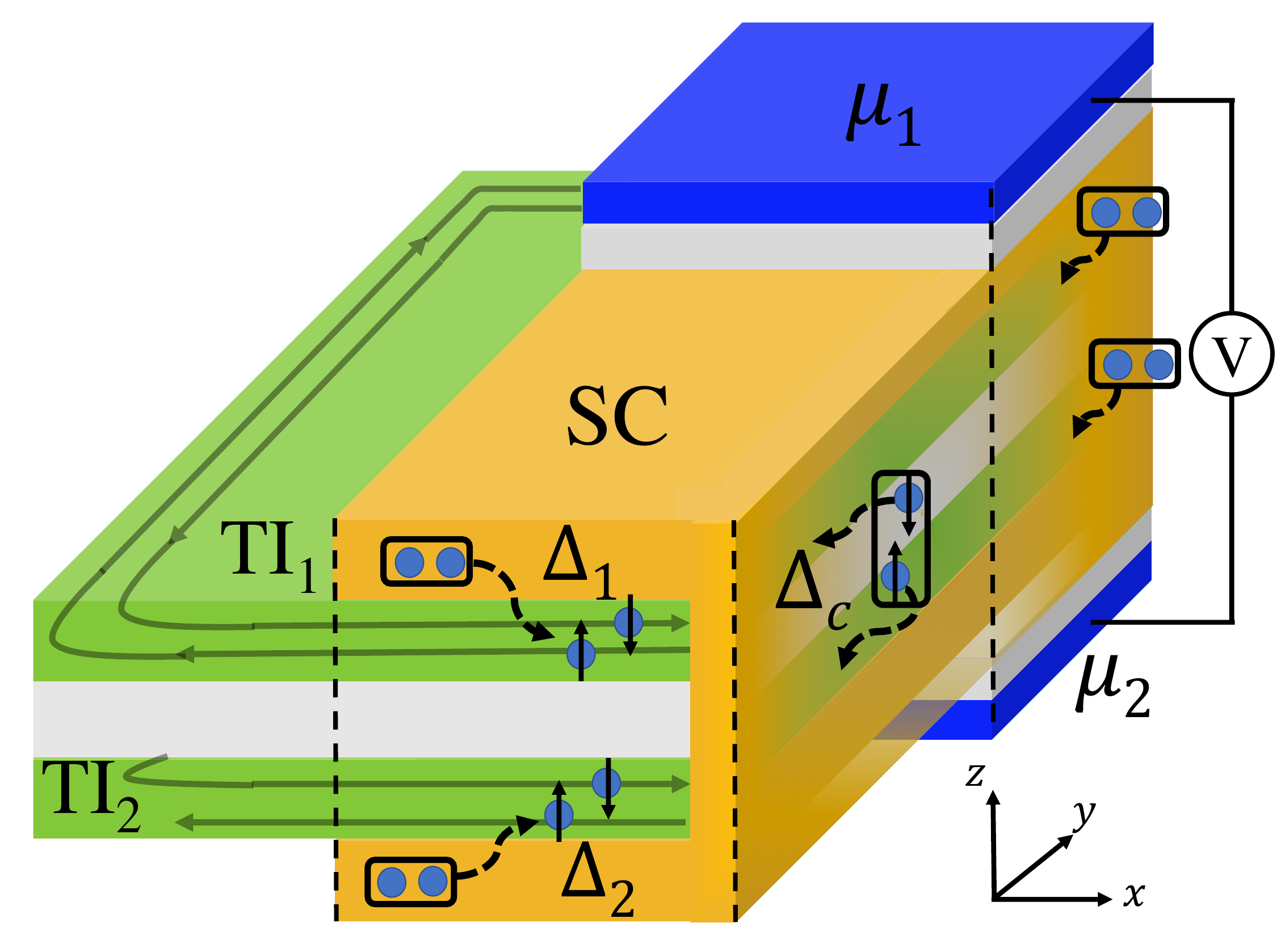}
\caption{Schematic of the bilayer hybrid system. It shows a bilayer consisting of two
identical TIs (green slabs). The helical edge states (with same helicity) of both TIs are in proximity to an $s$-wave superconductor (orange, grounded) and subject partly to an inter-layer gate voltage $V=\mu_1-\mu_2$ to control the chemical potentials of the two TI layers. Intra-layer ($\Delta_{1,2}$) as well as inter-layer ($\Delta_c$) pairings are possible.}
\label{fig.:fullsetup}
\end{figure}

We study a hybrid quantum system consisting of two identical HgTe quantum wells\cite{recher2012} in the QSH regime coupled to an $s$-wave superconductor (see Fig.~\ref{fig.:fullsetup}). Additionally, we assume that an inter-layer gate voltage allows us to tune the relative electron densities between the layers. This is a difficult task in experiments because the SC screens the influence of the gates. Hence, the geometry of SC and charging gates have to be chosen in a clever way. We compare our effective low energy results to exact numerical data on a microscopic bulk lattice model of the setup which, in reciprocal space, is described by the Bogoliubov de Gennes (BdG) Hamiltonian
\begin{align}
H_k=\begin{pmatrix}H_k^{\text{BL}}&\hat{\Delta}\\ \hat{\Delta}^\dag&-\mathcal T H_k^{\text{BL}} \mathcal T^{-1}\end{pmatrix}=\begin{pmatrix}H_k^{\text{BL}}&\hat{\Delta}\\ \hat{\Delta}^\dag&- H_{k}^{\text{BL}}\end{pmatrix},
\label{eqn:totham}
\end{align}
where the matrix structure is in Nambu space, $\mathcal T$ denotes the time reversal operator, and the second equality follows from time-reversal symmetry (TRS). The bilayer Hamiltonian $H_k^{\text{BL}}$ describing two identical, tunnel-coupled HgTe quantum wells has the following form in the layer pseudo-spin space:
\begin{align} \label{HBL}
H_k^{\text{BL}}=\begin{pmatrix}H^{\text{BHZ}}_k-\mu_1&h^t_k\\h^t_k&H^{\text{BHZ}}_k-\mu_2\end{pmatrix},
\end{align}
where the inter-layer potentials $\mu_{1,2}$ are crucial for some of the physics discussed in this work. In Eq.~(\ref{HBL}), $H_k^{\text{BHZ}}$ is assumed to be the Bloch Hamiltonian of the standard Bernevig-Hughes-Zhang model\cite{bhz2006} on a square lattice with lattice constant $a_\text{x,y}=a=1$, i.e.
\begin{align}
H_k^{\text{BHZ}}=\begin{pmatrix}h_k&0\\0&h^*_{-k}\end{pmatrix},
\end{align}
where one of the time-reversal conjugate (Kramers) blocks is given by $h_k=A (\sin(k_x) \sigma_x-\sin(k_y)\sigma_y)-(B \sigma_z + D \sigma_0)(4-2\cos(k_x)-2\cos(k_y)) + m \sigma_z$, and $\sigma_i$ are the standard Pauli matrices in  orbital space. In the basis $(|E,\uparrow>,|H,\uparrow>,|E,\downarrow>|H,\downarrow>)$ the tunnelling Hamiltonian $h_k^t$ is diagonal in spin space. Its matrix structure in orbital space is for spin up given by
\begin{align}
h_k^{t\uparrow}=\frac{1}{2}\begin{pmatrix}t_E&\alpha (\sin(k_x)+i\sin(k_y))\\ \alpha (\sin(k_x)-i\sin(k_y))& t_H \end{pmatrix},
\label{eqn:tunnelham}
\end{align}
and from TRS we have $\left(h_k^{t\uparrow}\right)^*=h_{-k}^{t\downarrow}$. Finally, the pairing matrix $\hat{\Delta}$ has the following structure in the layer pseudo-spin space
\begin{align}
\hat{\Delta}=\begin{pmatrix}\Delta_1&\Delta_c\\ \Delta_c & \Delta_2 \end{pmatrix},
\label{eqn:pairingmatrix}
\end{align}
where $\Delta_c$ denotes the crossed Andreev (inter-layer) pairing\cite{recher2002,fisher2002,Sato2010,feinberg2000,recher2013,braggio2019} from Cooper pairs delocalized over the two layers while $\Delta_1$ and $\Delta_2$ denote (intra-layer) pairing within the layers $1$ and $2$, respectively.

If the superconducting pairings $\Delta_{1,2,\text{c}}$ as well as the inter-layer voltage $V$ are much smaller than the bulk insulating gap of the quantum wells, we expect an effective edge description along the lines of Ref.~\onlinecite{Klino2014} to be a good approximation. There, the insulating bulk of the setup is neglected, thus only considering superconducting pairings in the helical edge states of the QSH bilayer system. For simplicity, we neglect direct tunneling of electrons between the layers, which is reasonable for $\alpha/a<\Delta_{1,2,\text{c}}$, and because the tunnelling decays exponentially with the distance of the two samples.

\begin{figure}
\centering
\includegraphics[width=0.9\linewidth]{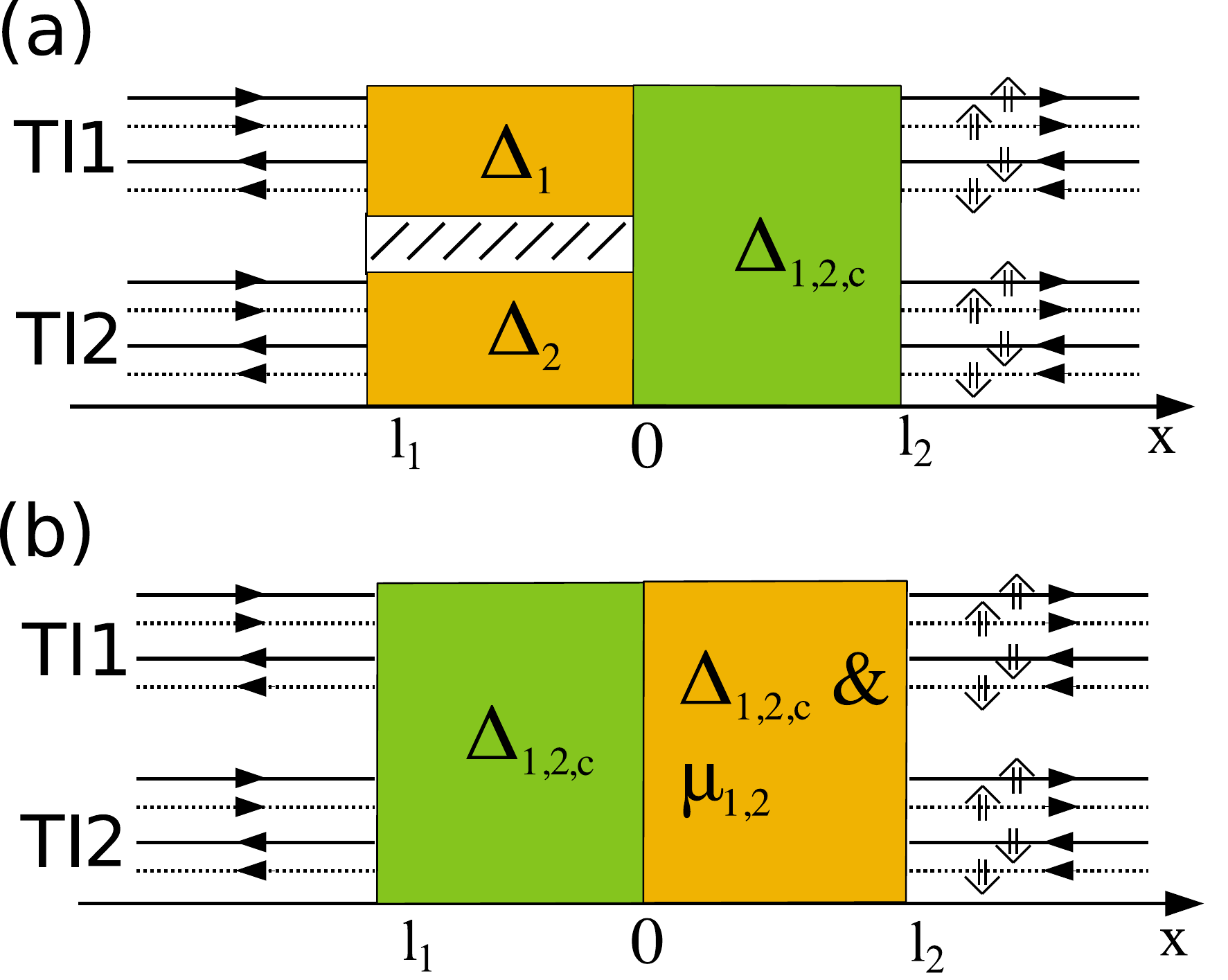}
\caption{N-S-S'-N junction of a bilayer topological insulator (TI1 and TI2). \textbf{(a)}: In region S $(l_1\leq x \leq 0)$ , we only consider intra-layer superconducting pairing $\Delta_{1,2}$, whereas in region S' $(0< x \leq l_2)$ we assume a dominating inter-layer pairing $\Delta_c$. \textbf{(b)}: In both superconducting regions S and S', we assume $\Delta_c^2>\Delta_1\Delta_2$. Here, the difference between regions S and  S' is a steplike variation of the relative density of the layers, i.e. a variation of $\mu_{1,2}$. The N sides are characterized by gapless helical edge states illustrated by arrows and spins in the figure. Full lines correspond to electron states and dashed lines to hole states (in a Bogoliubov-de Gennes picture).}
\label{fig.:setup}
\end{figure}

For the effective one-dimensional edge description along the $x$-axis, the edge modes are linearized around the Fermi-momentum $k_F$, and the fermion operators are expressed in terms of slowly varying left and right moving fields $\psi(x)=(R_1(x),L_1(x),R_2(x),L_2(x)$, $R_1^\dagger(x),L_1^\dagger(x),R_2^\dagger(x),L_2^\dagger(x))^T$. The resulting non interacting Hamiltonian can be written as
\begin{equation}\label{hamilton-2q}
\hat{H}=\frac{1}{2}\int dx \psi(x)^\dagger\mathcal{H}\psi(x)\textit{,}
\end{equation}
with the Hamiltonian density
\begin{eqnarray}\label{hamilton-density}
&&\mathcal{H}=\hbar v_\text{F} \hat{k}(\zeta_0\tau_0 s_3)-\frac{\mu_1}{2}\zeta_3\left(1+\tau_3\right)s_0-\frac{\mu_2}{2}\zeta_3\left(1-\tau_3\right)s_0 \nonumber \\
&&-\frac{\Delta_1}{2}\zeta_2\left(1+\tau_3\right)s_2-\frac{\Delta_2}{2}\zeta_2\left(1-\tau_3\right)s_2-\Delta_c\zeta_2\tau_1s_2\textit{,}
\end{eqnarray}\\
where the Pauli matrices $s_i$ ($\zeta_i$) [$\tau_i$] act on spin (electron-hole) [upper-lower QSH system] space and $\hat{k}=-i\hbar\partial_x$ is the momentum operator in real space representation. By the symmetric choice of $\mu_{1,2} = \pm V/2$, we parametrize the appearance of the chemical potentials in Eq.~(\ref{hamilton-density}) with a inter-layer gate voltage $V$ corresponding to the chemical potential difference $\mu_1-\mu_2$. We allow the superconducting pairing potentials $\Delta_{1,2}(x)$, $\Delta_c(x)$, and the chemical potential offsets $\mu_{1,2}(x)$ to depend on the spatial coordinate $x$. In the following, we put $\hbar = v_F = 1$.

Interestingly, our model refers to a superconducting hybrid system that preserves TRS. Hence, it can be classified by a topological quantum number $\mathbb{N}$ that is completely determined by the Fermi-surface properties, more precisely by the sign of the order parameter at the Fermi-surfaces in 1D \cite{zhang2010}. We employ this type of classification in the next section.

The characterization and detection of MBS are the main tasks of this article. It turns out that transport properties of hybrid bilayer structures are particularly suitable for the identification of MBS. Hence, we investigate below multi-terminal setups in which we assume that (i) the two layers of the bilayer QSH system are coupled to separate electron reservoirs and (ii) all superconducting regions are grounded. In Fig.~\ref{fig.:setup}, we schematically illustrate two setups of that kind that will be further analyzed below. In particular, both figures illustrate N-S-S'-N junctions in which the N regions are characterized by gapless helical edge states (with the same helicity in the two layers, for concreteness) and the S/S' regions are superconducting regions in which either different pairing amplitudes dominate or different inter-layer gate voltages are assumed.

\section{Spectroscopic properties}\label{sec.:sectionIII}
In this section, we discuss the spectroscopic features of our model. We identify BdG band inversions that can be generated by a variation of $\Delta_c$ with respect to $\Delta_{1,2}$ (at fixed $\mu_{1,2}$) or by a variation of $V=\mu_1-\mu_2$ with respect to $\Delta_c$ (at fixed $\Delta_c > \Delta_{1,2}$). Interestingly, Majorana bound states emerge in hybrid structures of QSH bilayers at interfaces between regions in space with different band orderings. Before we analyze such interfaces and the resulting bound states, we discuss different processes of tuning the band ordering of the BdG excitation spectrum in our system. The general idea is to identify possibilities of gap closings and reopenings as a function of adjustable parameters of the model. Some of the possibilities have been realized before (presented in Sec.~\ref{sec.:fall1})\cite{Klino2014}, others are not yet mentioned in the literature (presented in Sec.~\ref{sec.:fall2}).

\subsection{Gap closing by tuning pairing potentials}\label{sec.:sectionIIIa}
\label{sec.:fall1}

\begin{figure}
\centering
\includegraphics[width=1\linewidth]{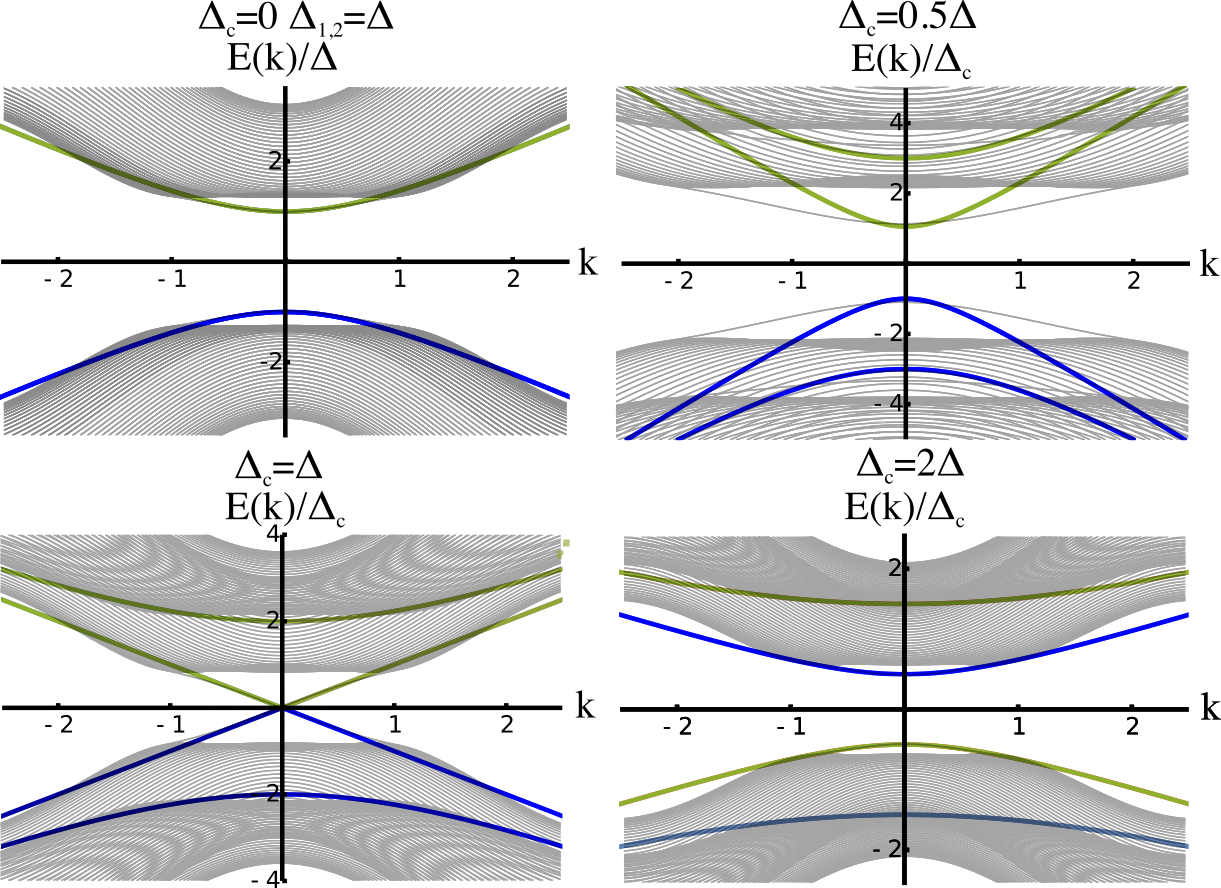}
\caption{ (color online) Illustration of the effective dispersion relation according to Eq.~(\ref{eq.:EnergieSCCROSS}) (thick green and blue lines) combined with the spectrum of an infinite (in y-direction) ribbon of width $N_x=50$ sites of a double HgTe quantum well in the QSH phase, where we set the nonzero parameters $A/\Delta=B/\Delta=m/\Delta=1.0$, while we assumed $\Delta_1=\Delta_2=\Delta$. The pairing terms $\Delta_{1,2}$ open a conventional superconducting gap, while its counter term $\Delta_c$ competes with them which can result in a gap closing. The gap closing happens at $\Delta_c^2=\Delta_1\Delta_2$. The color change of the bands (green vs. blue) after the reopening of the gap signals a band inversion.}
\label{fig.:DispersionWithoutV}
\end{figure}

First, we consider the effective edge Hamiltonian of Eq.~(\ref{hamilton-density}) under the choice $\mu_{1,2}=0$, i.e. in the absence of an inter-layer voltage. We look at the translational-invariant case in which the operator $\hat{k}$ can be replaced by the wave vector $k$ in Eq.~(\ref{hamilton-density}). Furthermore, we assume (for simplicity) that all spatially homogeneous pairing terms are purely real and positive.  Under this choice of parameters, the energy eigenvalues of the BdG Hamiltonian are given by
\begin{eqnarray}\label{eq.:EnergieSCCROSS}
E &=& \pm\frac{1}{\sqrt{2}}\Bigl( \Delta_1^2+\Delta_2^2+2\Delta_c^2+2 k^2 \\
&\pm& (\Delta_1+\Delta_2)\sqrt{(\Delta_1-\Delta_2)^2+4\Delta_c^2}\Bigr)^{1/2} \nonumber\; .
\end{eqnarray}
We note that each energy level has a twofold (Kramers) degeneracy because of TRS. To identify the points in parameter space, where the band ordering is changing, we first consider the gap closing conditions of the spectrum. In this model, the gap closing happens at $k=0$. Therefore, it is sufficient to look at the energy difference $\Delta E (k=0)$ of the lowest "conduction" band and the highest energy state of the "valence" band, i.e.
~\DIFaddend \begin{equation}\label{eq.gapclosing}
\Delta E (k=0)=\Delta_1+\Delta_2-\sqrt{(\Delta_1-\Delta_2)^2+4\Delta_c^2}\text{.}
\end{equation}
It is easy to see that the gap closes at $\Delta_c^2=\Delta_1\Delta_2$. The dispersion relation for different choices of system parameters is shown by the thick blue and green lines in Fig.~\ref{fig.:DispersionWithoutV}.
Evidently, a band inversion can be generated by a relative change of $\Delta_c$ with respect to $\Delta_{1,2}$.

The TRS ensures the supercondcuting pairings to be real and without loss of generality, we assume them to be positiv. By taking $\Delta_1=\Delta_2=\Delta$, we can identify two relevant pairing potentials at the Fermi-surface, namely $\Delta_+=\Delta+\Delta_c$ and $\Delta_-=\Delta-\Delta_c$. Accordingly, we are able to calculate a topological invariant for this superconductor that preserves TRS \cite{zhang2010}, i.e.
\begin{equation}
\mathbb{N}=\text{sgn}(\Delta_+)\text{sgn}(\Delta_-)=\text{sgn}(\Delta+\Delta_c)\text{sgn}(\Delta-\Delta_c)\text{,}
\end{equation}
where sgn($\Delta_+)$=sgn($\Delta+\Delta_c)$ is always greater than zero. Hence, for $\Delta_c>\Delta$, the invariant multiplies to $\mathbb{N}=-1$. In such hybrid structures, a Kramers pair of Majorana bound states appears at the boundary of two regions in space with a different topological invariant $\mathbb{N}$ as we demonstrate below.

We have corroborated our results based on the effective edge Hamiltonian of Eq.~(\ref{hamilton-density}) by numerical simulations on the full 2D lattice model specified in Eq.~(\ref{eqn:totham}) accounting for possible effects of the finite bulk energy gap on the edge physics. Our results fully confirm the gap closing and gap reopening behaviour from the competition of $\Delta_c$ and $\Delta_1,\Delta_2$. The numerical simulations are shown by the transparent lines in Fig.~\ref{fig.:DispersionWithoutV}.

As one of the main results of our present work, we show next that an inter-layer voltage $V$ allows us to switch between the different SC pairings thus enabling the flexible tunability of domain walls and MBS bound to them.

\subsection{Gap closing by tuning inter-layer voltage}\label{sec.:sectionIIIb}
\label{sec.:fall2}

\begin{figure}
\centering
\includegraphics[width=1\linewidth]{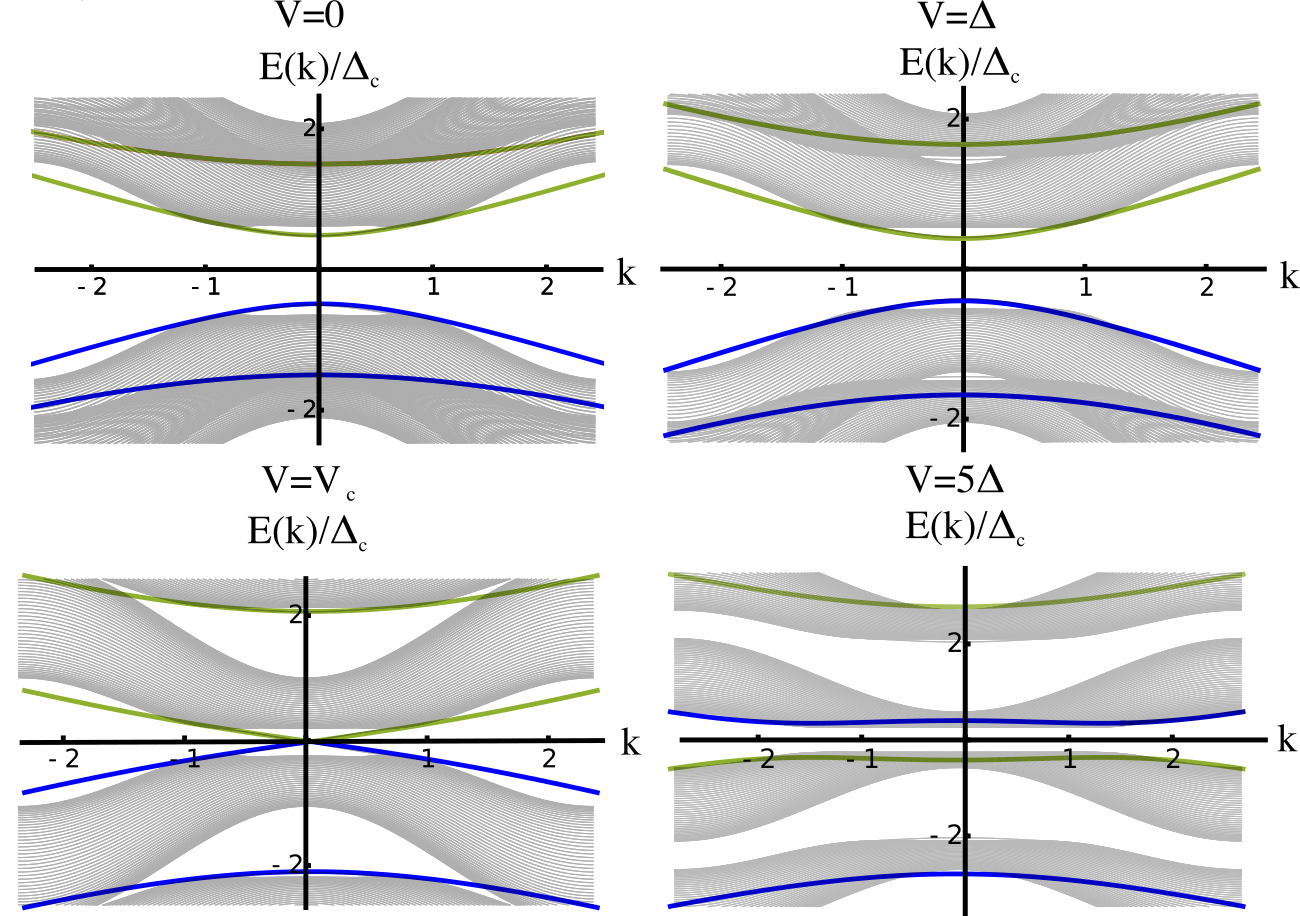}
\caption{(color online) Illustration of the dispersion relation according to the effective model in Eq.~(\ref{eq.:DispersionVoltage}) in combination with ribbon spectra of double-wells with SC pairings and varying strength of the inter-layer gate voltage. For illustrative reasons, we chose the nonzero parameters $A/\Delta=B/\Delta=m/\Delta=1.0$. The inter-layer voltage varies from $V=0.0,\Delta,2\sqrt{\Delta_c^2-\Delta^2} ,5\Delta$ as shown atop the panels. In all plots, we assume $\Delta_\text{c}=2\Delta$. We can realize the gap closing at $V_c=\pm2\sqrt{\Delta_c^2-\Delta^2}$ and its reopening for $|V|=|\mu_1-\mu_2|>|V_c|$. We illustrate the band inversion also by the colors of the bands.}
\label{fig.:Dispersion}
\end{figure}

In this section, we demonstrate that a gap closing can also be driven by a voltage, where we use the variation of the relative chemical potential parameters ($V=\mu_1-\mu_2$, where $\mu_{1,2} = \pm V/2$) of the two layers. On the basis of a spatially homogeneous version of the Hamiltonian stated in Eq.~(\ref{hamilton-density}) with the simplification $\Delta_1=\Delta_2\equiv\Delta$, we obtain the dispersion relation
\begin{eqnarray}\label{eq.:DispersionVoltage}
&&E=\frac{\pm 1}{2} \Bigl[ 4( \Delta^2+ \Delta_c^2)+4 k^2+V^2 \nonumber
\\
&&\pm4\sqrt{4 \Delta^2 \Delta_c^2+V^2 ( \Delta_c^2 + k^2)}\textit{ }\Bigr]^{1/2}\textit{,}\nonumber \\
\end{eqnarray}
where $V=\mu_1 - \mu_2$. The gap closing condition can be derived by solving Eq.~(\ref{eq.:DispersionVoltage}) for $k$ at zero energy, resulting in the solutions
\begin{eqnarray}
k_{E=0} &=& \frac{\pm i}{2}\big(2\Delta \pm\sqrt{4\Delta_c^2-V^2}\big)  \textit{.}
\label{eq.:WavevektorSCCROSSV}
\end{eqnarray}
To find the true gap closing condition, we have to identify (within this set of possible solutions) the ones that are purely real. Proper solutions are found for two choices of inter-layer voltage, i.e.
\begin{equation}
V_c=\pm 2 \sqrt{\Delta_c^2-\Delta ^2} \; .
\label{eq.:mu_cond}
\end{equation}
We coin the values of $V$ that fulfil Eq.~(\ref{eq.:mu_cond}) the critical $V_\text{c}$. They correspond to gap closings at wave vectors equal to zero. The choice of a non symmetric inter-layer voltage leads to a closing at finite $k$.

The dispersion relation for different values of the transition parameters $V$ is shown in Fig.~\ref{fig.:Dispersion}, where we assume $\Delta_c^2>\Delta^2$. Evidently, a band inversion also happens under this choice of the variation of parameters.
Again, we verify our results by numerical calculations on the full 2D lattice model given by Eq.~(\ref{eqn:totham}). The transparent lines in Fig.~\ref{fig.:Dispersion} complement with numerical data from microscopic simulations of this voltage driven transition the low energy model for various values of the inter-layer voltage $V$ and confirm the voltage induced gap closing at the transition to the trivial superconducting phase with increasing inter-layer voltage $V$.

In the presence of $V$, the BdG dispersion relation allows us to define two effective pairings at the Fermi surface, i.e.
\begin{eqnarray}
&&\tilde{\Delta}_+=\frac{1}{2}\sqrt{V^2+4\Delta^2}+\Delta_c \text{,}\nonumber \\
&&\tilde{\Delta}_-=\frac{1}{2}\sqrt{V^2+4\Delta^2}-\Delta_c \nonumber \text{,}
\end{eqnarray}
where $\tilde{\Delta}_+$ is greater than zero, because $V$ is chosen to be real. Hence, the topological invariant simplifies to
\begin{eqnarray}
\mathbb{N}=\text{sgn}(\tilde{\Delta}_-)=\text{sgn}(\frac{1}{2}\sqrt{V^2+4\Delta^2}-\Delta_c)\text{.}
\end{eqnarray}
Evidently, the sign of the invariant changes at $V=V_c$. By interfacing domains with $V<V_c$ and $V>V_c$, we again find Kramers pairs of MBS that are exponentially localized at the domain walls with spatially oscillating tails. We elaborate on these bound states in the following section.



\section{Majorana bound states}\label{sec.:sectionIV}

In this section, we discuss the emergence of Kramers pairs of MBS at the interface of the two S-S' junctions in Fig.~\ref{fig.:setup} (in the two layers). These bound states appear at junctions between sectors with different band orderings. As described in the previous section, there are two possibilities to achieve this task: (i) by a variation of $\Delta_c$ with respect to $\Delta_{1,2}$ and (ii) by a variation of $V=\mu_1-\mu_2$ at fixed $\Delta_c > \Delta_{1,2}$. We present results for both cases below. Technically, we look at a scattering problem of a N-S-S'-N junction in both cases. Since our system under consideration is time-reversal symmetric, we always obtain Kramers pairs of MBS. However, in our formalism, we employ scattering theory to identify a fingerprint of a MBS in a given scattering state. This approach typically yields one of the two Kramers partners for a particular choice of the scattering state. To identify the other Kramers partner, we need to look at the time-reversed scattering state in which direction of motion and spin are reversed.  The corresponding scattering states are specified in the next section and App.~\ref{sec.:appA}. The resulting bound states are then found by solving the secular equation that derives from the continuity conditions of the spinors and identifying solutions at zero excitation energy.

\subsection{MBS by variation of crossed pairing}\label{sec.:sectionIVa}

\begin{figure}
\centering
\includegraphics[width=1 \linewidth]{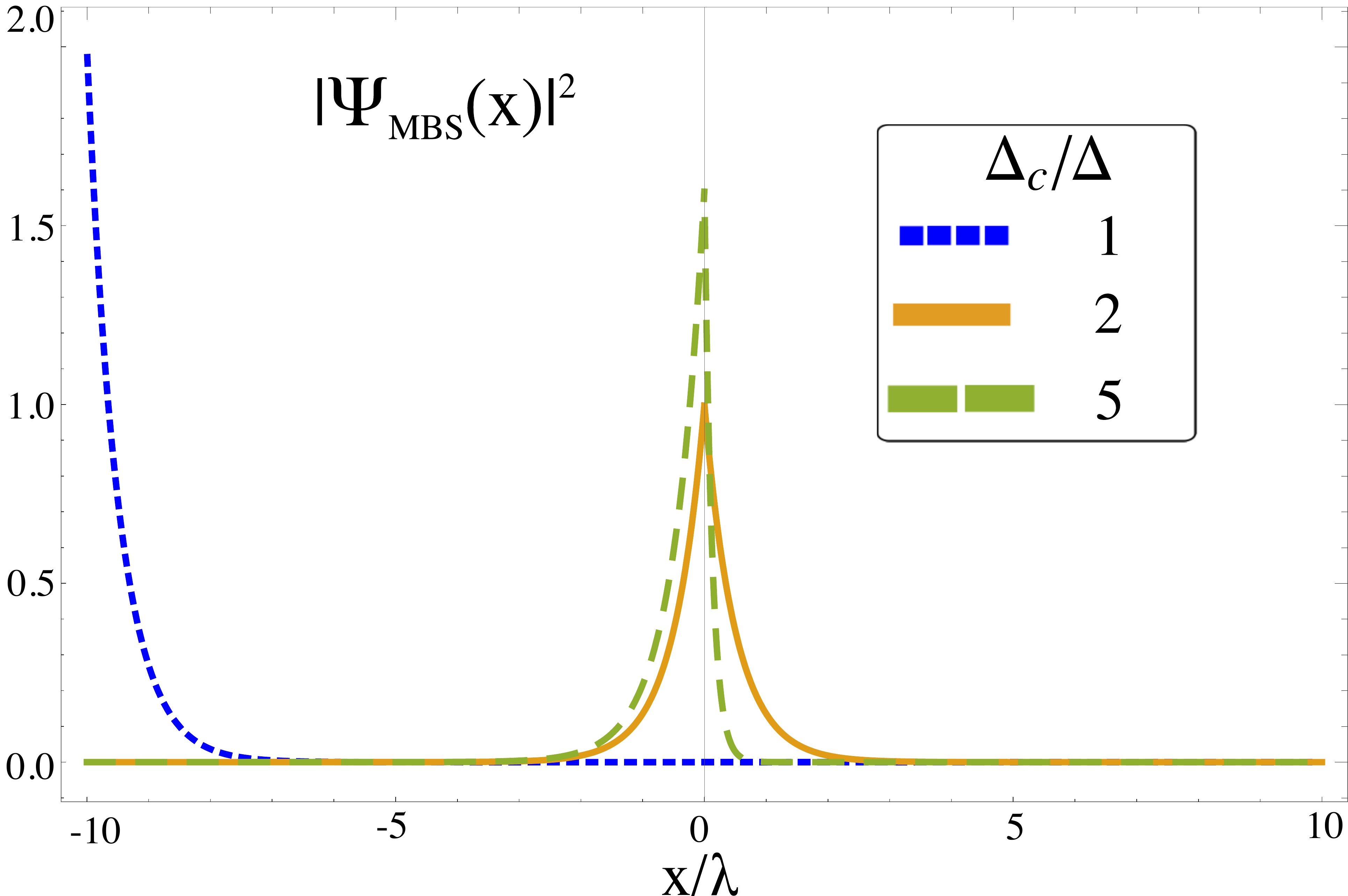}
\caption{Normalized absolute square of a bound state at zero excitation energy (i.e. a MBS) at the interface between S and S' in Fig.~\ref{fig.:setup} (a) for three different values of $\Delta_\text{c}$. Evidently, the MBS emerges for $\Delta_\text{c}^2 > \Delta^2$. For the particular choice $\Delta_c=2\Delta$, we observe a perfectly symmetric MBS because then the magnitudes of the gaps in the regions S and S' are equal. The parameters are chosen such that $\Delta_1=\Delta_2=\Delta=1/\lambda$, $l_1=-10\lambda$, and $l_2=10\lambda$. The realistic size of $\lambda$ varies from hundreds of nanometers to micrometers.}
\label{fig.:MBS-SC-SCCROSS}
\end{figure}

The N-S-S'-N junction under consideration is naturally divided into four regions in space which we name I-IV (from left to right). At each interface a continuity condition of the wave function has to be fulfilled because our Hamiltonian describes a set of first-order differential equations. On the basis of the coordinate system defined in Fig.~\ref{fig.:setup} (a), we label the spinor $\psi (x)$ accordingly $\psi_\text{I}(x)-\psi_\text{IV}(x)$ and take into account three different continuity conditions, i.e.
\begin{eqnarray} \label{eq:conti}
\psi_\text{I}(l_1)&=\psi_{\text{II}}(l_1)\textit{,}\nonumber \\
\psi_{\text{II}}(0)&=\psi_{\text{III}}(0)\textit{,} \\
\psi_{\text{III}}(l_2)&=\psi_{\text{IV}}(l_2)\textit{.}\nonumber
\end{eqnarray}
In region I and IV, we consider free propagating helical edge states in the bilayer setup in the absence of superconducting pairing. In region II, $\Delta_1=\Delta_2=\Delta$ is finite but $\Delta_c=0$. In region III, we tune the band inversion by $\Delta_c$ with respect to $\Delta_1=\Delta_2=\Delta$. All calculations are done for a scattering state with an incoming electron approaching the junction from region I in TI1. Similar bound states can be determined with other choices of scattering states. For better illustration, we only show the bound states in region II and III. In Fig.~\ref{fig.:MBS-SC-SCCROSS}, we plot the absolute square of the wave function of a solution to the secular equation at zero energy for different choices of $\Delta_c$ in region III with respect to finite $\Delta_1=\Delta_2=\Delta$. As expected, the MBS show an exponential decay with a localization length inversely proportional to the induced superconducting gaps ($\lambda_\text{MBS}\propto 1/\Delta E(k=0)$) in both regions, respectively. Realistic values for a superconducting localization length are discussed in the Conclusion.

\subsection{MBS by variation of inter-layer voltage}\label{sec.:sectionIVb}

\begin{figure}
\centering
\includegraphics[width=1\linewidth]{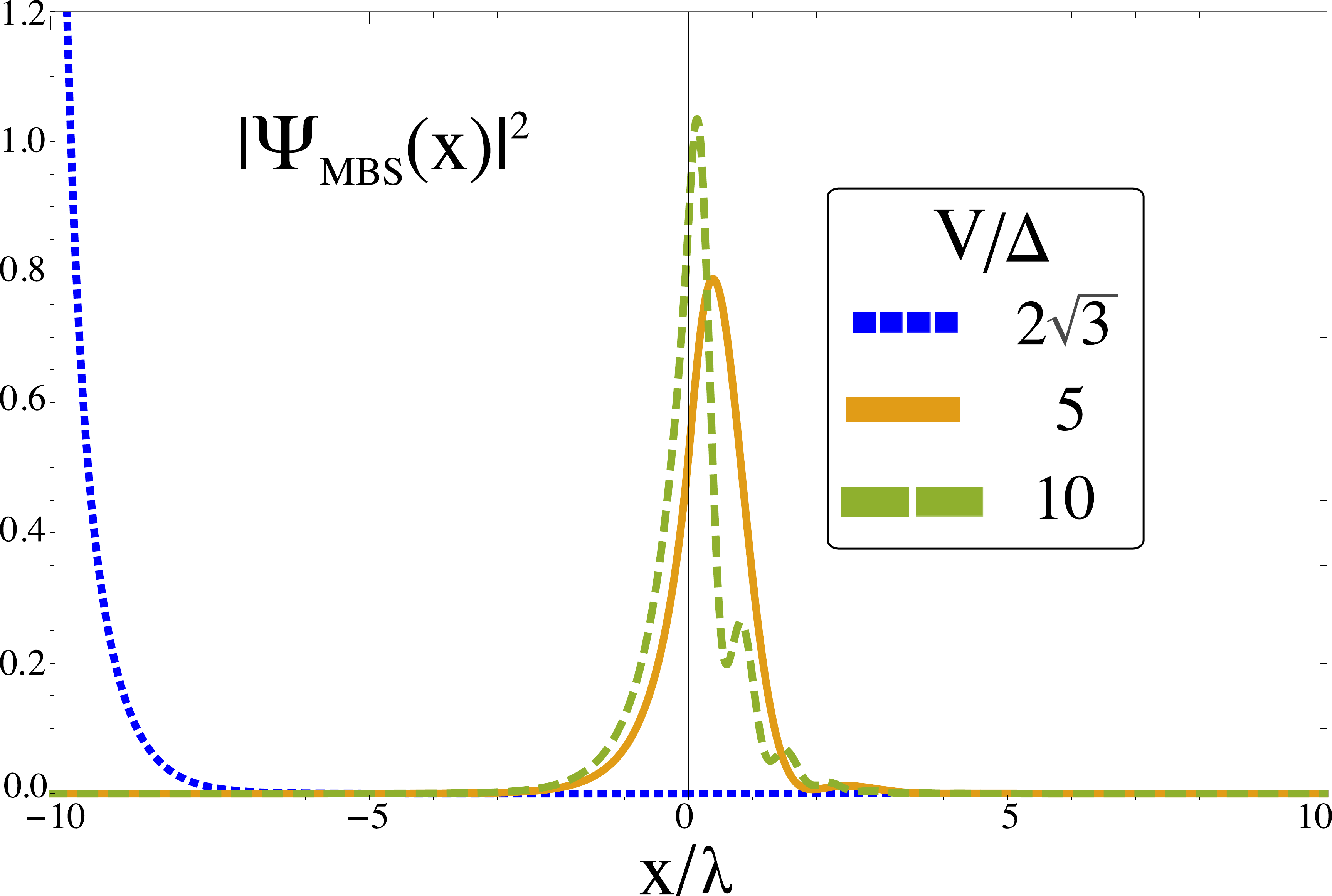}
\caption{Normalized absolute square of a bound state at zero excitation energy (i.e. a MBS) at the interface between S and S' illustrated in Fig.~\ref{fig.:setup} (b). Both superconducting regions have the same choice of pairing $\Delta_c = 2\Delta$ but they differ by the choice of inter-layer voltage $V=\mu_1-\mu_2$. In region II, we put $V=0$ whereas, in region III, we vary $V$ from $2\sqrt{3} \Delta$ to $10\Delta $. Further parameter choices are $\Delta_1 = \Delta_2 = \Delta=1/\lambda$, $l_1=-10\lambda$, and $l_2=10\lambda$. Typical parameter settings are discussed in the conclusion.\cite{molenkamp2014,kouwen2015,yacoby2017,molenkamp2017,Deacon2017}}
\label{fig.:MBS-SC-SCCROSS2}
\end{figure}

A similar analysis can be done for the setup shown in Fig.~\ref{fig.:setup} (b). Again, we divide the N-S-S'-N junction into four regions in space labeled I-IV. Similar to the previous section, we investigate a scattering event with an incoming electron approaching the junction from region I in TI1. Now, the superconducting pairing potentials are chosen homogeneously but a step-like variation of the inter-layer voltage discriminates region II from region III. For simplicity, we put $V=0$ in all regions in space and just vary $V$ in region III. The resulting bound states are plotted in Fig.~\ref{fig.:MBS-SC-SCCROSS2}. They emerge for $V>V_{\text{c}}$ in accordance to the band inversion properties discussed in Sec.~\ref{sec.:fall2}. Interestingly, the bound states shown in Fig.~\ref{fig.:MBS-SC-SCCROSS2} slightly differ from their counterparts in Fig.~\ref{fig.:MBS-SC-SCCROSS}. They are not centered precisely at $x=0$ but rather shifted to the right hand-side. Additionally, they develop oscillations (as a function of space) for larger steps in the inter-layer voltage. The oscillations arise only for $V> 2\Delta_c$ and their strength is independent of the normal pairing term $\Delta$. Since the two edges are connected only by the crossed Andreev pairing, a symmetric shift of $\mu_{1,2}$ closes the gap at $k=0$. For $V> 2\Delta_c$, the Fermi-momentum is away from $k=0$, which leads to a real (oscillating) part in the momentum (see Eq.~\ref{eq.:WavevektorSCCROSSV}). This can be seen in analogy to a spin orbit coupling induced gap opening for a one dimensional nanowire.\cite{lossklino2012} For more details, we refer to App.~\ref{sec.:appB}.
\section{Transport properties}\label{sec.:sectionV}
In this section, we aim to identify signatures of MBS that are observable in transport measurements. Similar to the previous section, we look at different types of N-S-S'-N junctions and calculate the corresponding scattering problem. Importantly, helicity puts tight constraints on the allowed scattering processes. For an incoming electron from region I in TI1 we have to take into account eight different scattering processes. In Tab.~\ref{tab.:processes}, we list them all and state which ones are allowed/forbidden by helicity. The allowed scattering processes are given by specific names: Electron cotunneling (EC), crossed electron cotunneling (CEC), local Andreev reflection (LAR), and crossed Andreev reflection (CAR). We stress that the reflection process of CAR is due to the one-dimensional system shown in Fig.~\ref{fig.:setup}. For the transport picture related to Fig.~\ref{fig.:fullsetup} one can see this process also as  transmission of a hole to TI2.
\begin{table}
\centering
\begin{tabular}{|l|l|l|}
\hline
\textbf{scattering process} & \textbf{SC} & \textbf{HC} \\ \hline
electron reflection in TI1 & $r_{e1}$ & f \\ \hline
electron reflection in TI2 & $r_{e2}$ & f \\ \hline
electron transmission in TI1 (EC) & $t_{e1}$ & a \\ \hline
electron transmission in TI2 (CEC) & $t_{e2}$ & a \\ \hline
Andreev reflection in TI1 (LAR) & $r_{h1}$ & a \\ \hline
Andreev reflection in TI2 (CAR) & $r_{h2}$ & a \\ \hline
hole transmission in TI1 & $t_{h1}$ & f \\ \hline
hole transmission in TI2 & $t_{h2}$ & f \\ \hline
\end{tabular}
\caption{List of allowed (a) and forbidden (f) scattering processes for a helical bilayer hybrid system assuming an incoming electron from the lhs in TI1. The first column defines the process, the second column the corresponding scattering coefficient (\textbf{SC}), and the last column the helicity constraint (\textbf{HC}) of this process.}
\label{tab.:processes}
\end{table}

The complete scattering states are composed of four terms (for the four different regions). They take the explicit forms
\begin{eqnarray}\label{eq.:wefu}
\psi_\text{I}(x) && =   \vec{\mathbf{e_1}}e^{ik_e^1x}+r_{e1}\vec{\mathbf{e_2}}e^{-ik_e^1x}+r_{e2}\vec{\mathbf{e_4}}e^{-ik_e^2x}\nonumber\\
&&+r_{h1}\vec{\mathbf{e_6}}e^{-ik_h^1x}+r_{h2}\vec{\mathbf{e_8}}e^{-ik_h^2x}\text{,}\nonumber \\
\psi_{\text{II}}(x)  &&=\sum_{i=1}^8a_i\vec{\mathbf{u}_i}e^{ik_ix}\text{,}\nonumber \\
\psi_{\text{III}}(x)  &&= \sum_{i=1}^8b_i\vec{\mathbf{v}_i}e^{ik'_ix}\text{,}\nonumber \\
 \psi_{\text{IV}}(x) &&=t_{e1}\vec{\mathbf{e_1}}e^{ik_e^1x}+t_{e2}\vec{\mathbf{e_3}}e^{ik_e^2x}\nonumber \\
&&+t_{h1}\vec{\mathbf{e_5}}e^{ik_h^1x}+t_{h2}\vec{\mathbf{e_7}}e^{ik_h^2x} \text{,}
\end{eqnarray}
where the momenta in the different regions in space are stated in App.~\ref{sec.:appA}.
In Eq.~(\ref{eq.:wefu}), the vectors $\vec{\mathbf{e_i}}$ in the two N regions are 8-dimensional Euclidean basis vectors. The vectors $\vec{\mathbf{u}_i}$ and $\vec{\mathbf{v}_i}$ that appear in the scattering states for the regions S and S' are too cumbersome to spell them out explicitly. We specify them in App.~\ref{sec.:appA}. The continuity conditions stated in Eq.~(\ref{eq:conti}) have to hold again. This set of equations allows us to calculate all coefficients of the scattering states (Eq.~\ref{eq.:wefu}), in particular, the transmission and reflection coefficients, $t_i$ and $r_i$, respectively.

The spin helical leads in our setup provide a stringent selectiveness in scattering processes listed in Tab.~\ref{tab.:processes}. In a five-terminal setup, in which the left N region and the right N region are each separately contacted by two electron reservoirs in each TI and the regions S and S' are grounded, we investigate the (local and non-local) differential conductance $dI_i/dV_j$. The indices $i$ and $j$ run from 1 to 4 (over the numbering of the four electron reservoirs connected to TI1 and TI2). The current $I_i$ corresponds to the current that leaves or enters reservoir $i$ and the voltage $V_j$ corresponds to a bias applied locally to reservoir $j$.

Let us, for concreteness, specify the allowed scattering processes for an incoming right-moving electron in TI1: (i) Electron cotunnelling (EC) from reservoir 1 (connected to the lhs of TI1) to reservoir 3 (connected to the rhs of TI1). This process is reflected by a finite net current $G_{\text{EC}}^{31}(V_1-V_3)$ between reservoirs 1 and 3; (ii) crossed electron cotunnelling (CEC) from reservoir 1 to reservoir 4 (connected to the rhs of TI2). This process is reflected by a finite net current $G_{\text{CEC}}^{41}(V_1-V_4)$ between reservoirs 1 and 4; (iii) local Andreev reflection (LAR) at reservoir 1 with corresponding current $2G_{\text{LAR}}V_1$; and (iii) crossed Andreev reflection (CAR) between reservoirs 1 and reservoir 2 (connected to the lhs of TI2) with corresponding current $G_{\text{CAR}}^{21}(V_1+V_2)$~\cite{Falci}.

If we continue the thought for any possible incoming particle at any desired reservoir, the full multi-terminal current-voltage relation can be written in compact form as
\begin{equation}
\begin{pmatrix}
I_1\\
I_2\\
I_3\\
I_4
\end{pmatrix}=
\begin{bmatrix}
G_1' & G_{\text{CAR}}^{12} & -G_{\text{EC}}^{13} & -G_{\text{CEC}}^{14} \\
G_{\text{CAR}}^{21} & G_2' & -G_{\text{CEC}}^{23} & -G_{\text{EC}}^{24}\\
-G_{\text{EC}}^{31} &- G_{\text{CEC}}^{32} & G_3' & G_{\text{CAR}}^{34}\\
-G_{\text{CEC}}^{41} & -G_{\text{EC}}^{42} & G_{\text{CAR}}^{43} & G_4'
\end{bmatrix}
\begin{pmatrix}
V_1\\
V_2\\
V_3\\
V_4
\end{pmatrix}\textit{,}
\end{equation}
where the diagonal terms of the conductance matrix are given by
\begin{eqnarray}
G_1'=&2G_{\text{LAR}}+G_{\text{CAR}}^{21}+G_{\text{EC}}^{31}+G_{\text{CEC}}^{41}\nonumber \\
G_2'=&2G_{\text{LAR}}+G_{\text{CAR}}^{12}+G_{\text{EC}}^{42}+G_{\text{CEC}}^{32}\nonumber \\
G_3'=&2G_{\text{LAR}}+G_{\text{CAR}}^{43}+G_{\text{EC}}^{13}+G_{\text{CEC}}^{23}\nonumber \\
G_4'=&2G_{\text{LAR}}+G_{\text{CAR}}^{34}+G_{\text{EC}}^{24}+G_{\text{CEC}}^{14}\textit{.}
\end{eqnarray}
Along the lines of BTK \cite{BTK}, we can relate the non-linear conductance $dI_i/dV_j$ of our hybrid system (for simplicity, at zero temperature $T=0$) to corresponding reflection and transmission probabilities through
\begin{eqnarray}\label{eq.:conductivity1}
&&\mathbf{\frac{dI}{dV}}= \frac{e^2}{h}\\
&&\begin{bmatrix}
G_1(eV_1) & R_{\text{CAR}}^{12}(eV_2) & -T_{\text{EC}}^{13}(eV_3) & -T_{\text{CEC}}^{14}(eV_4) \\
R_{\text{CAR}}^{21}(eV_1) & G_2(eV_2) & -T_{\text{CEC}}^{23}(eV_3) & -T_{\text{EC}}^{24}(eV_4)\\
-T_{\text{EC}}^{31}(eV_1) & -T_{\text{CEC}}^{32}(eV_2) & G_3(eV_3) & R_{\text{CAR}}^{34}(eV_4)\\
-T_{\text{CEC}}^{41}(eV_1) & -T_{\text{EC}}^{42}(eV_2) & R_{\text{CAR}}^{43}(eV_3) & G_4(eV_4)
\end{bmatrix}\nonumber
\end{eqnarray}
with $R=|r|^2$, $T=|t|^2$, and the diagonal terms
\begin{eqnarray}\label{eq.:conductivity}
G_1=&1+R_{\text{LAR}}(eV_1)\nonumber \\
G_2=&1+R_{\text{LAR}}(eV_2)\nonumber \\
G_3=&1+R_{\text{LAR}}(eV_3)\nonumber \\
G_4=&1+R_{\text{LAR}}(eV_4)\textit{.}
\end{eqnarray}
Equation~(\ref{eq.:conductivity1}) nicely illustrates the beauty of the helical bilayer system. The five-terminal setup allows us to measure individually the transmission/reflection probabilities of each scattering process. This remarkable feature solely stems from the particular constraints due to helicity summarized in Tab.~\ref{tab.:processes}. In the following sections, we argue that this particular advantage of our system (compared to non-helical analogues) gives rise to unique transport signatures that can be related to the presence/absence of (Kramers partners of) MBS at the S-S' interface.

\subsection{Transport in the absence of inter-layer voltage}\label{sec.:sectionVa}

Since MBS are bound states at zero excitation energy $E=0$, we first focus on the corresponding scattering coefficients. In particular, we try to identify zero-bias anomalies in the structure illustrated in Fig.~\ref{fig.:setup} (a). Fortunately, it is possible to derive and display analytical results for all relevant transmission/reflection probabilities. Under the parameter choice $\mu_1=\mu_2=0$ and $\Delta_1=\Delta_2=\Delta$ and $l_1<0<l_2$, we obtain
\begin{eqnarray}\label{eq.:LAR}
R_{\text{LAR}}^{\text{E}=0}=&\frac{\left(e^{4 \Delta  l_1}-e^{4 \Delta  l_2}\right)^2}{N^2} \text{,}\\ \nonumber \\
R_{\text{CAR}}^{\text{E}=0}=&\frac{4 e^{4 \Delta  (l_1+l_2)} \sinh ^2(2 \Delta_c l_2)}{N^2}\text{,}\\ \nonumber \\ \label{eq.:CAR}
T_{\text{EC}}^{\text{E}=0}=&\frac{16 e^{4 \Delta  (l_1+l_2)} \cosh ^2(\Delta_c l_2) \cosh ^2(\Delta  (l_1-l_2))}{N^2}\text{,} \\ \nonumber \\ \label{eq.:EC}
T_{\text{CEC}}^{\text{E}=0}=&\frac{16 e^{4 \Delta  (l_1+l_2)} \sinh ^2(\Delta_c l_2) \sinh ^2(\Delta  (l_1-l_2))}{N^2}\text{} \label{eq.:CEC}
\end{eqnarray}
with $N=\left(e^{2 \Delta l_1}+e^{2 l_2 (\Delta -\Delta_c)}\right) \left(e^{2 \Delta  l_1}+e^{2 l_2 (\Delta +\Delta_c)}\right)$. Let us first interpret these results before we discuss concrete signatures thereof in the differential conductance. Evidently, for $\Delta_\text{c}=0$, the probabilities $R_{\text{CAR}}$ and $T_{\text{CEC}}$ vanish because the two QSH layers are totally disconnected in this case. Then, regions II and III are identical which is the reason why the finite probabilities $R_{\text{LAR}}$ and $T_{\text{EC}}$ follow $\tanh^2\big((|l_1|+l_2)\Delta\big)$ and $1-\tanh^2\big((|l_1|+l_2)\Delta\big)$, respectively. For increasing superconducting region, LAR is getting more dominant and EC is suppressed.

\begin{figure}
\centering
\includegraphics[width=1\linewidth]{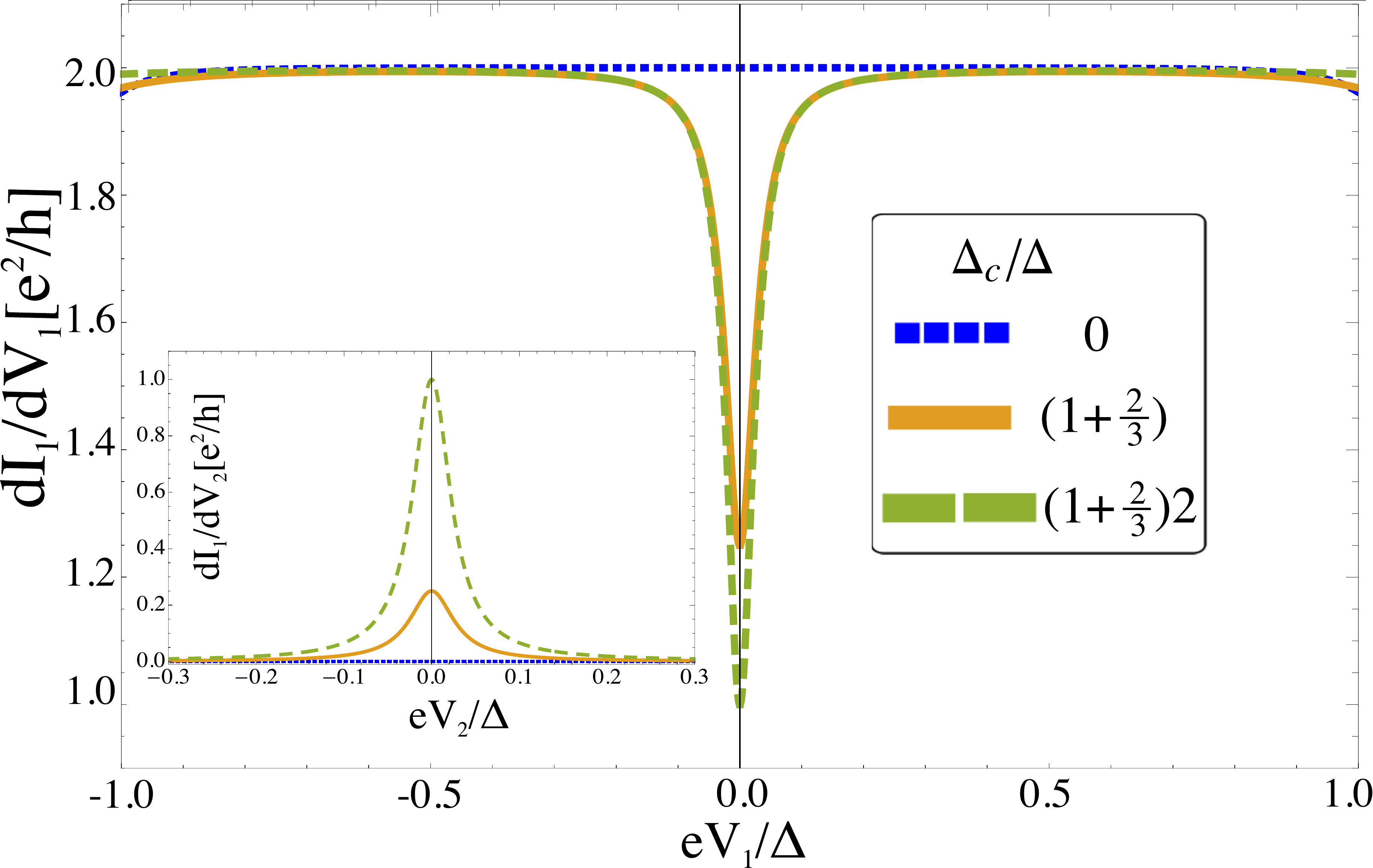}
\caption{Local (non-local) differential conductance $dI_1/dV_1$ ($dI_1/dV_2$) as function of $eV_1/\Delta$ ($eV_2/\Delta$) for different values of $\Delta_\text{c}$. A zero bias dip (peak) arise exactly at the {\it transport phase transition} (see text) where $\Delta_c=(1+\frac{|l_1|}{l_2})\Delta$. For higher values of $\Delta_c$ both differential conductances reach the values of $e^2/h$. We chose the parameters such that $\Delta_1=\Delta_2=\Delta$, $l_1=-2\lambda$ and $l_2=3\lambda$. }
\label{fig.:dIdV11ohneVoltage}
\end{figure}

More interesting physics arises, of course, for finite $\Delta_\text{c}$. Remarkably, the length scales $l_1$ and $l_2$ enter into the transport coefficients in a non-trivial way. By a comparison of the exponentials in the transport coefficients in Eqs.~(\ref{eq.:LAR})-(\ref{eq.:CEC}), we identify a new condition for a transport phase transition, given by $\Delta_c=(1+\frac{|l_1|}{l_2})\Delta$. At this point, all transport processes at $E=0$ have the same probability $1/4$.
For larger values of $\Delta_\text{c}$, we find vanishing LAR, EC, CEC, and a perfect CAR at $E=0$. Therefore, we see that this perfect CAR is in one-to-one relation with the presence of MBS in our system. The resulting local ($dI_1/dV_1$) and non-local ($dI_1/dV_2$) differential conductances as a function of biases $V_1$ and $V_2$, respectively, are shown in Fig.~\ref{fig.:dIdV11ohneVoltage}.

Evidently, the pronounced zero bias dip (peak) in the local (non-local) differential conductance can serve as a clear signature of the presence of Kramers pairs of MBS in our system. However, it is interesting to realize that for transport properties the condition for zero bias features is slightly altered compared to the spectroscopic properties discussed in the previous section. This peculiarity has been realized in simpler hybrid structures based on single-layer quantum spin Hall systems before~\cite{Fleckenstein}. While for the spectroscopic appearance of zero energy bound states, the criterion $\Delta_\text{c}>\Delta$ is relevant (for the choice $\Delta_1=\Delta_2=\Delta$), the zero bias features in transport only emerge if $\Delta_\text{c}>(1+\frac{|l_1|}{l_2})\Delta$. In the latter case, the different weights of spatial inhomogeneities matter with respect to the length scales $l_1$ and $l_2$.

\subsection{Transport in the presence of inter-layer voltage}\label{sec.:sectionVb}

\begin{figure}
\centering
\includegraphics[width=1\linewidth]{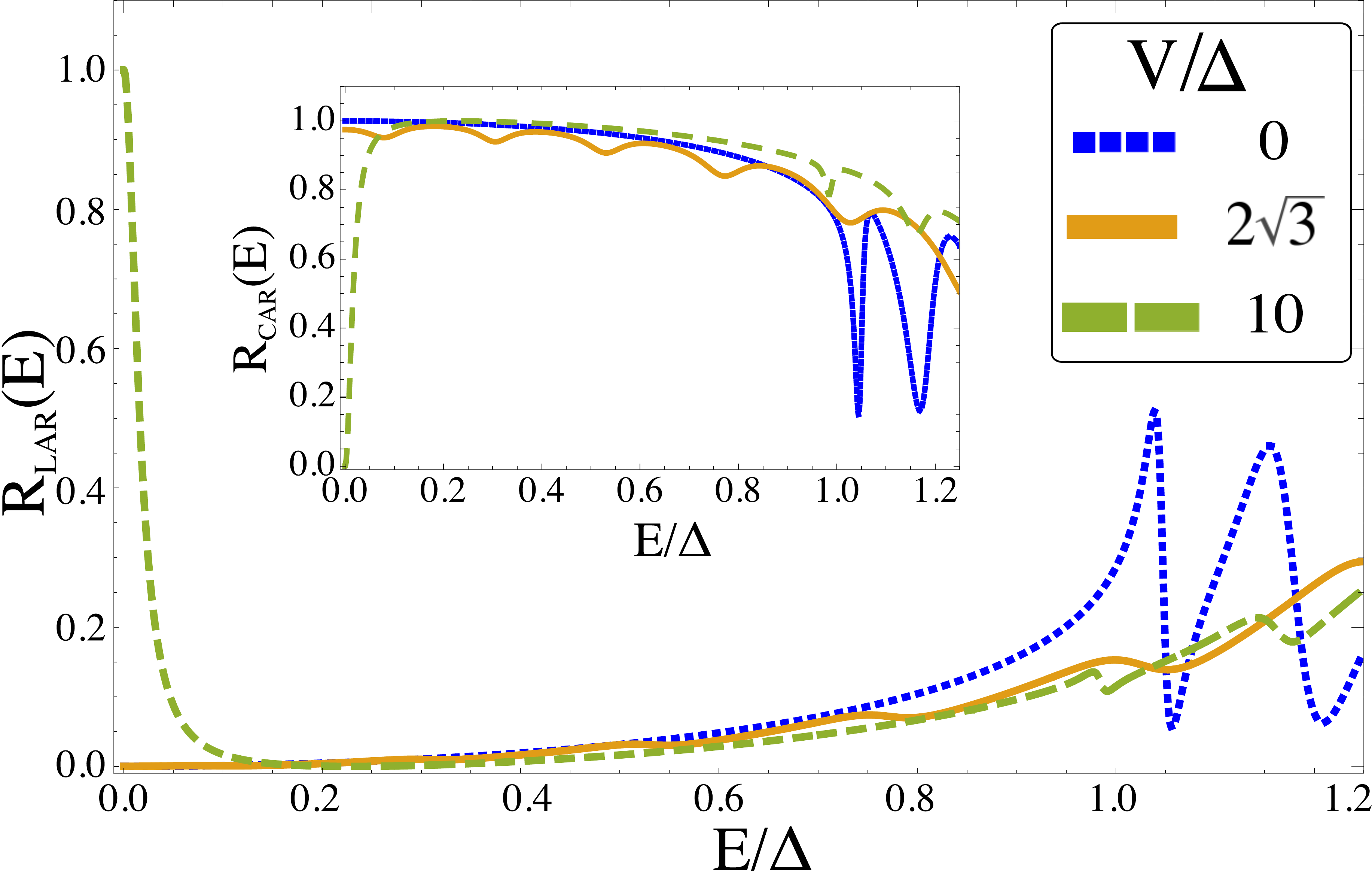}
\caption{Scattering probabilities for different values of $V$. We clearly observe the emergence of zero bias peaks or dips (in $R_\text{LAR}$ and $R_\text{CAR}$) for inter-layer voltages characterized by $V>V_\text{c}$, while the cotunneling probabilities ($T_\text{EC}$ and $T_\text{CEC}$) are negligible within the subgap regime. We set the other parameters to $\Delta_1=\Delta_2=\Delta$, $\Delta_c=2\Delta$, $l_1=-2\lambda$ and $l_2=8\lambda$. Here, $V_c=2\sqrt{3}\Delta$.}
\label{fig.:CoefficientsmitV}
\end{figure}

For finite inter-layer voltage, our analytical results for the transport probabilities are too lengthy to be explicitly written down here. Therefore, we first graphically display the relevant reflection ($R_{\text{LAR}}$ and $R_{\text{CAR}}$) probabilities in Fig.~\ref{fig.:CoefficientsmitV} as a function of the excitation energy before we turn to the presentation of the corresponding differential conductance.

At the critical inter-layer voltage defined by $V_{\text{c}}$ via Eq.~(\ref{eq.:mu_cond}) we realize a qualitative change in the transmission/reflection properties. For $V <V_\text{c}$, at small excitation energies, transport is dominated by CAR because of our choice of pairing potentials $\Delta_c=2\Delta$ in Fig.~\ref{fig.:CoefficientsmitV}. This behavior drastically changes for $V > V_\text{c}$. Then, LAR dominates for $E/\Delta \approx 0$. This comes from the fact that the energetic distance between the Fermi-energies of the two layers ($\mu_{1,2}$) is greater than the gap opened by the crossed pairing. The oscillations at larger excitation energies $E/\Delta \geq 1$ are Fabry-P\'erot resonances due to quasiparticle interference. The width of the zero bias peaks or dips can be varied by the system size. The larger the superconducting regions, the smaller is the width of the feature at zero excitation energy.

\begin{figure}
\centering
\includegraphics[width=1\linewidth]{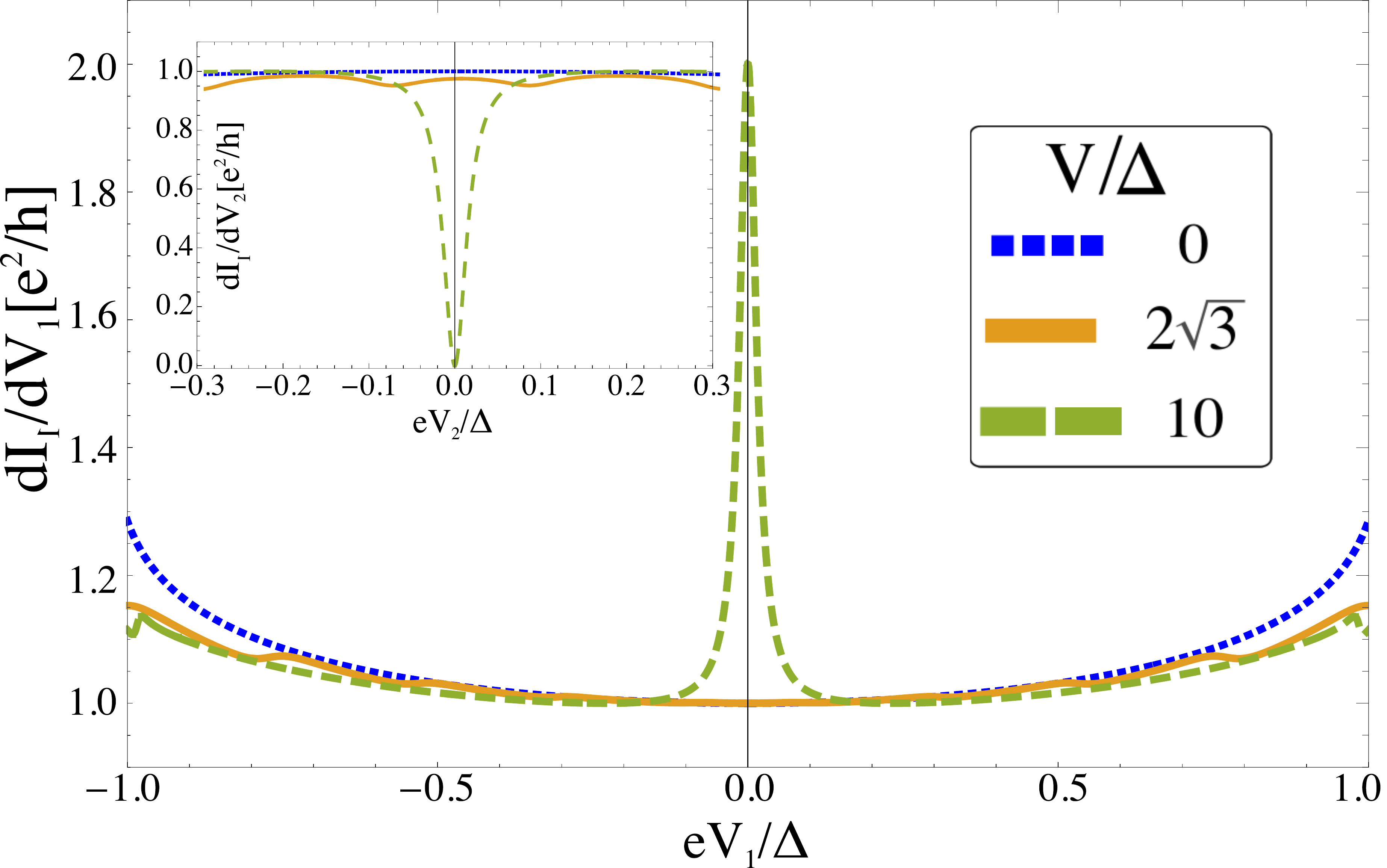}
\caption{Local (nonlocal) differential conductance $dI_1/dV_1$ ($dI_1/dV_2$) as function of $eV_1/\Delta$ ($eV_2/\Delta$) for different values of $V$ (in units of $\Delta$). A zero bias dip (peak) arises only for $V>V_c$ (green dashed line). In contrast to Fig.~\ref{fig.:dIdV11ohneVoltage}, we now find a dip instead of a peak at zero bias in the presence of MBS. We set the other parameters to $\Delta_1=\Delta_2=\Delta$, $\Delta_c=2\Delta$, $l_1=-2\lambda$, and $l_2=8\lambda$. Here, $V_c=2\sqrt{3}\Delta$.}
\label{fig.:dIdV11Voltage}
\end{figure}

The corresponding differential conductances are shown in Fig.~\ref{fig.:dIdV11Voltage}.
Interestingly, we can identify a dip in the non-local conductance $dI_1/dV_2$ in the regime of large inter-layer voltage in which a MBS emerges at the step in $V$. This dip should be put in contrast to the peak that we observe in Fig.~\ref{fig.:dIdV11ohneVoltage} where the MBS is tuned by a step-like variation of $\Delta_c$. Hence, we observe that zero-bias anomalies (peaks or dips) signal the presence of MBS in the five-terminal setup.

\subsection{Length dependence of the MBS features}\label{sec.:sectionVc}

\begin{figure}
\centering
\includegraphics[width=1\linewidth]{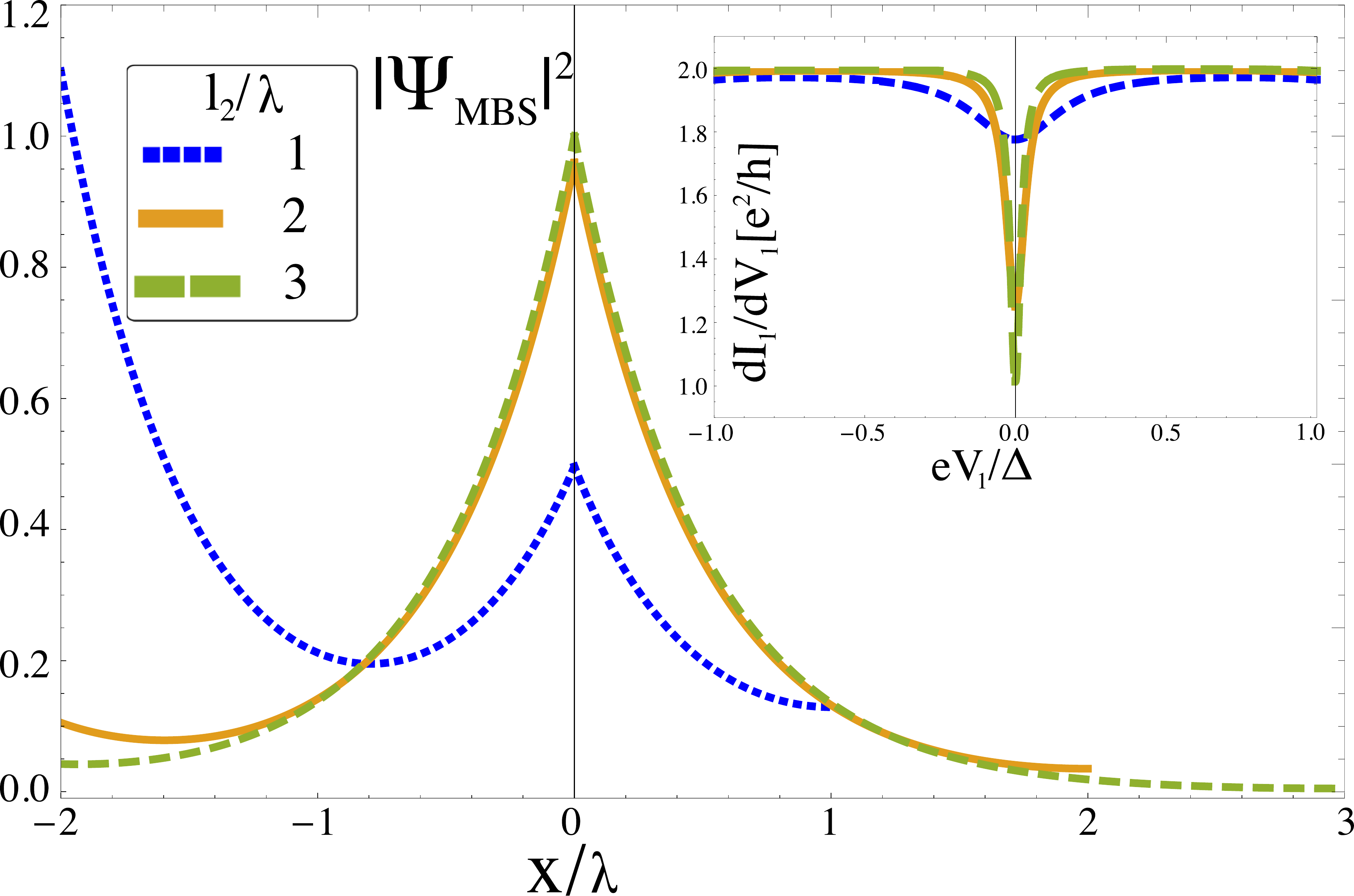}
\caption{Normalized absolute square of a bound state at zero energy at the interface in Fig.~\ref{fig.:setup} (a) for three different sizes of region III, while we assumed $\Delta_\text{c}=2\Delta$ for $\Delta_1=\Delta_2=\Delta$ and $V=0$. In addition we show the corresponding local conductance $dI_1/dV_1$. The bound state gets shifted to the left interface with decreasing length, while the conductance also vanishes. We used for illustrative reasons the size of region II $|l_1|=2\lambda$ and do not display the free propagating waves in the normal regions.}
\label{fig.:dIdVMBSlength}
\end{figure}

We have introduced in Sec.~\ref{sec.:sectionVa} an effective transport phase transition weighted by the length scales $l_1$ and $l_2$. Now we show that the length scales of region II ($|l_1|$) and the length of region III ($l_2$) can affect the MBS and its conductance features. Therefore, we take a parameter choice where the MBS and its signatures already exist and vary the length of region III. We show in Fig.~ \ref{fig.:dIdVMBSlength} the resulting local conductance with the related MBS for three sizes of region III.

Interestingly, we find that with the decrease of region III the dip of the conductance vanishes. Coincidently, the MBS gets shifted to the left interface, but still peaks at $x=0$. The reason for the length dependencies is that the scattering process of a particle at zero energy is not only influenced by the magnitude of the potential but also by the length of the acting potential. The new condition of the emergence of a zero bias anomaly on a N-S-S'-N junction is given by the strength of the order parameter times the corresponding length.

\section{Summary and Conclusion}\label{sec.:sectionVI}

In this article, we have compared the low energy model of a bilayer hybrid system of helical edge states with induced superconducting pairing to numerical data on a microscopic tight-binding model, confirming the validity of the effective model, and the competition between a direct pairing and a crossed pairing superconducting gap. Going beyond previous work on a similar setup, we have introduced an inter-layer voltage that can suppress the crossed pairing driving the system from the crossed pairing phase to the direct pairing phase by merely tuning the voltage. This feature represents a simple experimental knob for tuning the topological phase of the system. With the help of a  spatially varying inter-layer voltage, domains with different effective pairings hosting Kramers pairs of MBS are shown to be achievable. Additionally, we have discussed a five-terminal setup for a proposed transport experiment. We have shown that the emergence of the MBS is in accordance with zero bias anomalies in local and non-local differential conductance. The zero bias behaviour is characterized by peaks or dips in the differential conductance. Finally, we have also addressed the question of how the lengths of the different scattering regions can affect transport signatures and the emergence of Kramers pairs of MBS in distinct ways.

Let us conclude with a brief discussion of the feasibility of our proposal. We require that the distance $d$ of the two QSH layers should be smaller than the coherence length $\lambda$ of the superconductor (e.g. for Al $ \lambda \approx 1.6\mu$m). Moreover, inter-edge tunnelling should be avoided. In a previous work on bilayer QSH insulators, some of us have shown that inter-layer tunnelling almost vanishes for a barrier thickness of $d>10$nm \cite{recher2012}. Thus, there is a window of interlayer distance $d$ that is favorable for our idea.

We thank C.~Fleckenstein, F.~Keidel,  J.~Klinovaja,  D.~Loss, H.~W.~Schumacher and P.~Silvestrov for fruitful discussions. We acknowledge financial support by the DFG (SPP1666, SFB1170 ``ToCoTronics'', SFB1143 ``Correlated Magnetism: From Frustration to Topology'' ID 247310070 , AS 327/5-1, RE 2978/7-1 and the DACH project on ``Majorana and parafermions in topological insulators''), the ENB Graduate school on ``Topological Insulators'', and the W\"urzburg-Dresden Cluster of Excellence on Complexity and Topology in Quantum Matter (EXC 2147, project-id 39085490).

\newpage
\appendix
\section{MBS wave functions in the absence of an inter-layer gate voltage}\label{sec.:appA}
To get more insight into the shape of the MBS, we examine here the structure of the wave vectors in more detail. First, we take into account the setup shown in Fig.~\ref{fig.:setup} (a). Region I implies the free TI edge states of both samples with right and left moving electrons or holes following the dispersion $E=\pm k_{e,h}^{1,2}$ ($\hbar=v_\text{F}=1$). The related wave vectors for electrons and holes in region II, from $x=l_1$ ($l_1<0$) to $x=0$, with energies in the superconducting gap are given by
\begin{eqnarray}
k_{\pm}=\pm i\sqrt{\Delta ^2-E^2}\text{.}
\end{eqnarray}
We identify an exponential decay into region II with a decay length inversely proportional to the superconducting gap $\lambda_{II}= 1/\Delta$ at zero energy.\\
In region III from $x=0$ to $x=l_2$, where the crossed pairing enters the game, we obtain the following wave vectors:
\begin{eqnarray}\label{eq.:ZEwavevectorsCROSS}
k_{\pm,\pm}=\pm i  \sqrt{ (\Delta \pm  \Delta_c)^2-E^2}
\end{eqnarray}
Again, we identify an exponential decay into region III, inversely proportional to the gap $|\Delta-\Delta_c|$, illustrated in Fig.~\ref{fig.:DispersionWithoutV}. Since we consider the MBS, which has zero excitation energy we find a vanishing real part $\operatorname{Re} \left[k_{\pm}\right]=0$. This implies that there is no oscillating part in the wave function. The imaginary part of the wave vectors at zero energy vanishes for $\Delta_\text{c}=\Delta$, which is also an indication for the phase transition. The exponential decaying MBS in Fig.~\ref{fig.:MBS-SC-SCCROSS} has a perfect symmetric shape for $\Delta_c=2\Delta$, which comes from the fact, that in this case the energy gaps in both regions have the same magnitude and conclusively the same decay length.\\
Assuming a symmetric length of region II and III around $x=0$ ($l_1=-l_2=-L$) and employing continuity conditions, we find the zero energy wave function, up to a normalization factor:
\onecolumngrid
\begin{equation}
\psi_{\text{I}}(x)=
\begin{pmatrix}
1\\
r_{\text{e1}}\\
0\\
r_{\text{e2}} \\
0\\
r_{\text{h1}} \\
0\\
r_{\text{h2}}
\end{pmatrix}\text{ ,}
\end{equation}
\begin{equation}\label{eq.:WEFUII}
\psi_{\text{II}}(x)=
A\cdot\begin{pmatrix}
\cosh (2 \Delta_c L) \cosh (\Delta  \tilde{L}_1)+\cosh ( \Delta\tilde{L}_3)\\
0\\
-\sinh (2 \Delta_c L) \sinh (\Delta  \tilde{L}_1)\\
0\\
0\\
i [\cosh (2 \Delta_c L) \sinh (\Delta \tilde{L}_1)-\sinh ( \Delta\tilde{L}_3)]\\
0\\
-i \sinh (2 \Delta_c L) \cosh (\Delta \tilde{L}_1)
\end{pmatrix}\text{ ,}
\end{equation}
\\
\begin{eqnarray}\label{eq.:WEFUIII}
\psi_{\text{III}}(x)=
A&&\cdot\begin{pmatrix}
\frac{1}{2} [\cosh ( \Delta \tilde{L}_3+\Delta_c x)+\cosh (\Delta\tilde{L}_1+ \Delta_c \tilde{L}_2)+\cosh ( \Delta\tilde{L}_1 -\Delta_c \tilde{L}_2)+\cosh ( \Delta\tilde{L}_3-x\Delta_c)]\\
0\\
-[\sinh (\Delta_c x) \sinh ( \Delta \tilde{L}_3)+\sinh (\Delta  \tilde{L}_1) \sinh (\Delta_c\tilde{L}_2)]\\
0\\
0\\
i [ \cosh (\Delta_c\tilde{L}_2)\sinh (\Delta \tilde{L}_1)-\cosh (\Delta_c x) \sinh ( \Delta \tilde{L}_3)]\\
0\\
-i [\cosh (\Delta \tilde{L}_1) \sinh (\Delta_c \tilde{L}_2)-\sinh (\Delta_c x) \cosh (\Delta\tilde{L}_3)]
\end{pmatrix}\text{ ,}
\end{eqnarray}
\\
\begin{equation}
\psi_{\text{IV}}(x)=
\begin{pmatrix}
t_\text{e1} \\
0\\
t_\text{e2} \\
0\\
t_\text{h1} \\
0\\
t_\text{h2}\\
0
\end{pmatrix}\text{,}
\end{equation}
\twocolumngrid
\noindent
where $A=\frac{2 e^{2 L (2 \Delta +\Delta_c)}}{\left(e^{4 \Delta L}+e^{2 \Delta_c L}\right) \left(e^{2 L (2 \Delta +\Delta_c)}+1\right)}$, $\tilde{L}_1=L+x$, $\tilde{L}_2=2L-x$ and  $\tilde{L}_3=3L-x$.
The beauty of the scattering wavefunction approach can also be seen in Eqs.~(\ref{eq.:WEFUII}) and (\ref{eq.:WEFUIII}), by comparing the process related entries on both sides. \\
Assuming the incident electron is a right mover coming from TI1, which restricts us to only four allowed scattering states, mentioned in the main text. Those processes are related to the four non vanishing entries of the wavefunction. The resulting MBS arises at the interface at $x=0$. For simplicity, we only display scattering states in region II and region III $(\Psi_\text{II}(x)$ and $\Psi_\text{III}(x))$. The resulting bound state has not yet the properties of a MBS. To create the MBS we first write the bound state in the compact form
\begin{eqnarray}\label{eq.boundstate}
\Psi(x)&&=(\Theta (x+l_1)-\Theta (x))\Psi_\text{II}(x)\nonumber \\
&&+(\Theta (x)-\Theta (-x+l_2))\Psi_\text{III}(x)\text{.}
\end{eqnarray}
Next we superpose the bound state Eq.~(\ref{eq.boundstate}) with its charge conjugated partner. Note that the charge conjugation operator must satisfy the condition
\begin{equation}
\mathcal{C}\mathcal{H}_{BdG}\mathcal{C}^{-1}=U_c^\dagger\mathcal{H}_{BdG}^*U_c=-\mathcal{H}_{BdG}\text{,}
\end{equation}
which is fulfilled by $\mathcal{C}=U_c\mathcal{K}=\zeta_1\mathcal{K}$, where $\mathcal{K}$ denotes the complex conjugation and $U_c$ is a unitary transformation.  Conclusively, we can construct two MBS by the superpositions
\begin{eqnarray}
&&\Psi_{\text{MBS},1}(x)=\Psi(x)+\mathcal{C}\Psi(x)\text{,}\nonumber \\
&&\Psi_{\text{MBS},2}(x)=i(\Psi(x)-\mathcal{C}\Psi(x))\text{.}
\end{eqnarray}
It turns out, that the charge conjugated wave function $\mathcal{C}\Psi(x)$ is exactly the wave function for the bound state when we assume an incoming hole in TI1. Interestingly, we obtain another MBS which is related to a symmetry, which we call sample symmetry $\mathcal{S}=\tau_1$. This symmetry changes the sample space if it acts on the Hamiltonian, so that there are two additional MBSs, namely
\begin{eqnarray}
&&\Psi_{\text{MBS},3}(x)=\mathcal{S}\Psi_{\text{MBS},1}(x)\text{,}\nonumber \\
&&\Psi_{\text{MBS},4}(x)=\mathcal{S}\Psi_{\text{MBS},2}(x)\text{.}
\end{eqnarray}
These MBS correspond to bound states if we assume an incoming electron in TI2 with the superposition of its charge conjugated wave function. Those states are orthogonal to each other. Finally, we find with the TRS operator $\mathcal{T}=i s_2\mathcal{K}$, eight different MBS located at the interfaces. They can be specified as
 \begin{eqnarray}
 &&\Psi_{\text{MBS},1}(x)=\Psi(x)+\mathcal{C}\Psi(x)\text{,}\nonumber \\
&&\Psi_{\text{MBS},2}(x)=i(\Psi(x)-\mathcal{C}\Psi(x))\text{,}\nonumber \\
&&\Psi_{\text{MBS},3}(x)=\mathcal{S}(\Psi(x)+\mathcal{C}\Psi(x))\text{,}\nonumber \\
&&\Psi_{\text{MBS},4}(x)=\mathcal{S}i(\Psi(x)-\mathcal{C}\Psi(x))\text{,}\nonumber \\
 &&\Psi_{\text{MBS},5}(x)=\mathcal{T}(\Psi(x)+\mathcal{C}\Psi(x))\text{,}\nonumber \\
&&\Psi_{\text{MBS},6}(x)=\mathcal{T}i(\Psi(x)-\mathcal{C}\Psi(x))\text{,}\nonumber \\
&&\Psi_{\text{MBS},7}(x)=\mathcal{T}\mathcal{S}(\Psi(x)+\mathcal{C}\Psi(x))\text{,}\nonumber \\
&&\Psi_{\text{MBS},8}(x)=\mathcal{T}\mathcal{S}i(\Psi(x)-\mathcal{C}\Psi(x))\textit{.}
\end{eqnarray}

\section{MBS wave functions in the presence of an inter-layer gate voltage}\label{sec.:appB}

\begin{figure}[hhh]
\centering
\includegraphics[width=1.0\linewidth]{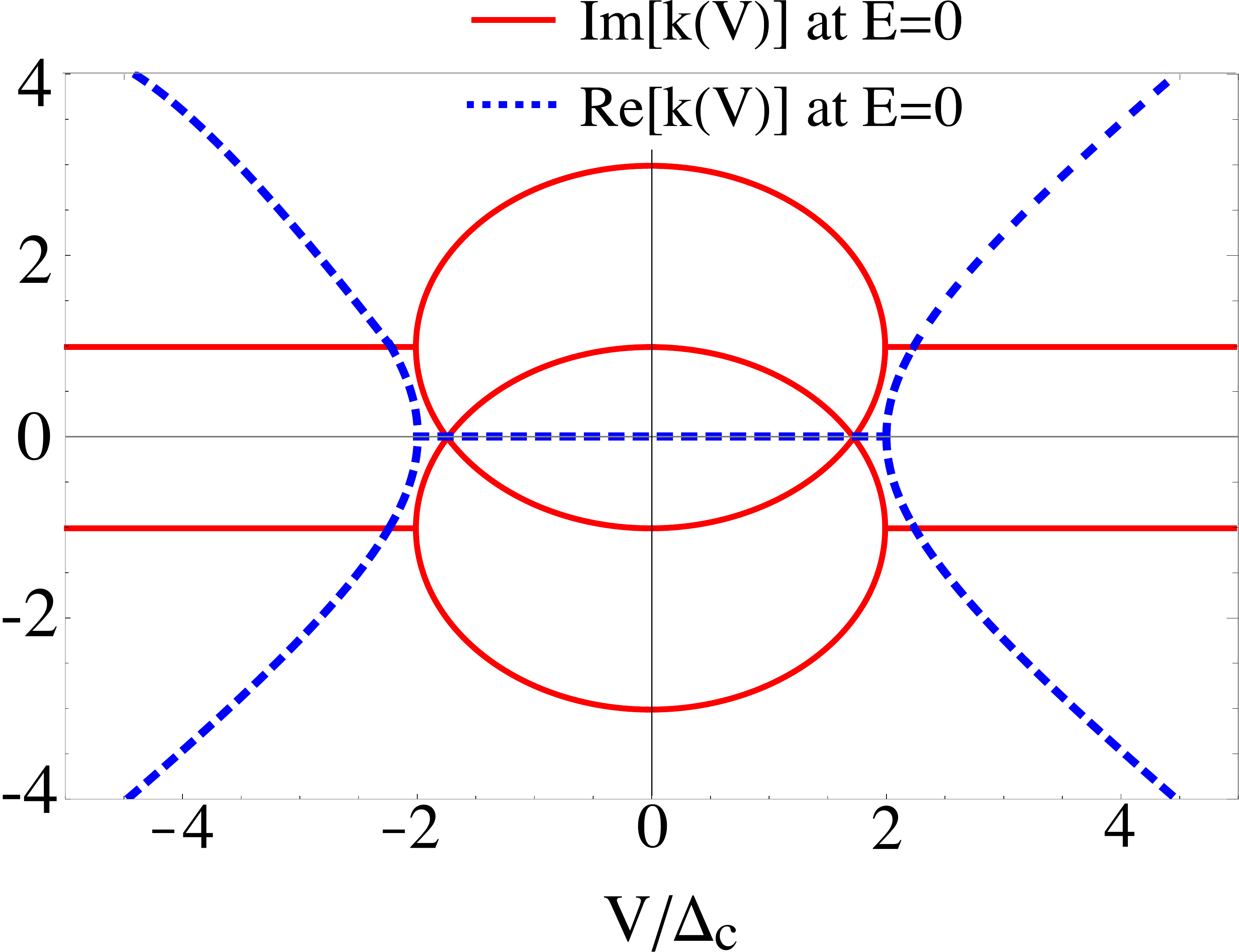}
\caption{Real part (blue, dashed line) of the wavevectors at $E=0$, in dependence of the symmetric inter-layer voltage $V$ in units of $\Delta_c=2\Delta$. One sees a splitting of the vectors at $V=2\Delta_\text{c}$.  Imaginary part (red, solid line) of the wavevectors at $E=0$. For values $V\geq2\Delta_\text{c}$ the imaginary part stays constant.}
\label{fig.:IMwavevectorMU}
\end{figure}

Here, we display the momenta for the region with the inter-layer gate voltage and examine the wave vectors for zero energy.
We use the setup in Fig.~\ref{fig.:setup} (b), where we assume $\Delta_c>\Delta$ in region II and an additional inter-layer gate voltage in region III. Again, region I and IV represent the free edge states in TI1 and TI2. The related wave vectors in region II are given by Eq.~(\ref{eq.:ZEwavevectorsCROSS}) and were already discussed in App.~\ref{sec.:appA}.
In region III, we vary the inter-layer gate voltage. The resulting momenta in Eq.~(\ref{eq.:wefu}) are given by
\begin{eqnarray}\label{eq.:wavevectorinterlayerbias}
k_{\pm,\pm}(E)=&&\frac{\pm1}{2} \bigg( 4E^2+V^2-4 \Delta ^2-4 \Delta_c^2  \nonumber \\
&&\pm 4 \sqrt{4 \Delta ^2 \Delta_c^2-V^2 \left(\Delta ^2-E^2 \right)} \bigg)^{1/2}\textit{,}
\end{eqnarray}
where the inter-layer gate voltage is defined as $V=\mu_1-\mu_2$. The creation of the scattering state for an incoming rightmoving electron in TI1 is done in the same manner as in App.~\ref{sec.:appA}. We get more information of the MBS in Fig.~\ref{fig.:MBS-SC-SCCROSS2}, by looking at the momenta in region III.
Fig.~\ref{fig.:IMwavevectorMU} shows the imaginary and real parts of Eq.~(\ref{eq.:wavevectorinterlayerbias}) at zero energy in dependence of $V/\Delta_c$ under the assumption that $\Delta_\text{c}=2\Delta$.
This result explains, why we find in region III an oscillating part, together with an exponential decay.

\end{document}